\documentstyle[onecolumn,psfig]{mn}

\newif\ifAMStwofonts

\ifoldfss
  \ifCUPmtlplainloaded \else
    \NewTextAlphabet{textbfit} {cmbxti10} {}
    \NewTextAlphabet{textbfss} {cmssbx10} {}
    \NewMathAlphabet{mathbfit} {cmbxti10} {} 
    \NewMathAlphabet{mathbfss} {cmssbx10} {} 
  \fi
  \ifAMStwofonts
    \ifCUPmtlplainloaded \else
      \NewSymbolFont{upmath} {eurm10}
      \NewSymbolFont{AMSa} {msam10}
      \NewMathSymbol{\upi}     {0}{upmath}{19}
      \NewMathSymbol{\umu}     {0}{upmath}{16}
      \NewMathSymbol{\upartial}{0}{upmath}{40}
      \NewMathSymbol{\leqslant}{3}{AMSa}{36}
      \NewMathSymbol{\geqslant}{3}{AMSa}{3E}

      \let\leq=\leqslant \let\le=\leqslant
      \let\geq=\geqslant \let\ge=\geqslant
    \fi
  \fi
\fi 

\ifnfssone
  \newmathalphabet{\mathit}
  \addtoversion{normal}{\mathit}{cmr}{m}{it}
  \addtoversion{bold}{\mathit}{cmr}{bx}{it}
  \newmathalphabet{\mathbfit} 
  \addtoversion{normal}{\mathbfit}{cmr}{bx}{it}
  \addtoversion{bold}{\mathbfit}{cmr}{bx}{it}
  \newmathalphabet{\mathbfss} 
  \addtoversion{normal}{\mathbfss}{cmss}{bx}{n}
  \addtoversion{bold}{\mathbfss}{cmss}{bx}{n}
  \ifAMStwofonts
    \ifCUPmtlplainloaded \else
      \UseAMStwoboldmath
      \makeatletter
      \new@mathgroup\upmath@group
      \define@mathgroup\mv@normal\upmath@group{eur}{m}{n}
      \define@mathgroup\mv@bold\upmath@group{eur}{b}{n}
      \edef\UPM{\hexnumber\upmath@group}
      \new@mathgroup\amsa@group
      \define@mathgroup\mv@normal\amsa@group{msa}{m}{n}
      \define@mathgroup\mv@bold\amsa@group{msa}{m}{n}
      \edef\AMSa{\hexnumber\amsa@group}
      \makeatother
      \mathchardef\upi="0\UPM19
      \mathchardef\umu="0\UPM16
      \mathchardef\upartial="0\UPM40
      \mathchardef\leqslant="3\AMSa36
      \mathchardef\geqslant="3\AMSa3E

      \let\leq=\leqslant \let\le=\leqslant
      \let\geq=\geqslant \let\ge=\geqslant
    \fi
  \fi
\fi 

\ifnfsstwo
  \DeclareMathAlphabet{\mathbfit}{OT1}{cmr}{bx}{it}
  \SetMathAlphabet\mathbfit{bold}{OT1}{cmr}{bx}{it}
  \DeclareMathAlphabet{\mathbfss}{OT1}{cmss}{bx}{n}
  \SetMathAlphabet\mathbfss{bold}{OT1}{cmss}{bx}{n}
  \ifAMStwofonts
    \ifCUPmtlplainloaded \else
      \DeclareSymbolFont{UPM}{U}{eur}{m}{n}
      \SetSymbolFont{UPM}{bold}{U}{eur}{b}{n}
      \DeclareSymbolFont{AMSa}{U}{msa}{m}{n}
      \DeclareMathSymbol{\upi}{0}{UPM}{"19}
      \DeclareMathSymbol{\umu}{0}{UPM}{"16}
      \DeclareMathSymbol{\upartial}{0}{UPM}{"40}
      \DeclareMathSymbol{\leqslant}{3}{AMSa}{"36}
      \DeclareMathSymbol{\geqslant}{3}{AMSa}{"3E}

      \let\leq=\leqslant \let\le=\leqslant
      \let\geq=\geqslant \let\ge=\geqslant
    \fi
  \fi
\fi 

\ifCUPmtlplainloaded \else
  \ifAMStwofonts \else 
    \def\upi{\pi}
    \def\umu{\mu}
    \def\upartial{\partial}
  \fi
\fi

\title[Millisecond and Binary Pulsars as Nature's Frequency Standards]
{Millisecond and Binary Pulsars as Nature's Frequency Standards. \\
{\LARGE II. Effects of Low-Frequency Timing Noise on Residuals and Measured 
Parameters}}
\author[Sergei M. Kopeikin]
       {Sergei M. Kopeikin \thanks{On leave from: ASC FIAN, 
       Leninskii Prospect, 53, Moscow 117924, Russia}\\ \\
      FSU Jena, TPI, Max-Wien-Platz 1, D - 07743, Jena, Germany}
\date{Accepted.............199... ;  
      Received ............199... ; 
      in original form ...........199... }
\pagerange{\pageref{firstpage}--\pageref{lastpage}}
\pubyear{1999}

\def\LaTeX{L\kern-.36em\raise.3ex\hbox{a}\kern-.15em
    T\kern-.1667em\lower.7ex\hbox{E}\kern-.125emX}

\begin{document}

\label{firstpage}

\maketitle

\begin{abstract}
Millisecond and binary pulsars are the most stable natural
frequency standards. They can be applied to a number of principal
problems of modern astronomy and time-keeping metrology including the 
search for stochastic gravitational
wave background in the early universe, testing General Relativity and
establishing of new ephemeris time scale. The full exploration of pulsar
properties requires 
obtaining proper unbiased estimates of the spin and orbital 
parameters, a problem which deserves a special investigation. These 
estimates depend essentially 
on the random noise component being revealed in the residuals of time of 
arrivals (TOA) and having a different physical origin.
In the present paper, the influence of low-frequency ("red") timing 
noise with 
spectral indices from 1 to 6 on TOA residuals, variances, and covariances of 
estimates of measured parameters of single and binary pulsars are studied. 
In order to
determine their functional dependence on time, an analytical
technique of processing of observational data in time domain is developed. Data
processing in time domain is more informative since it takes into account both
stationary component of noise and its non-stationary countepart. Data processing
in frequency domain is valid if and only if the noise is stationary. Our
analysis
includes a simplified timing model of a binary pulsar in a 
circular orbit and procedure of estimation of pulsar parameters and residuals
under the influence of red noise. We reconfirm, in accordance with results of
previous authors, that 
uncorrelated white noise of 
errors of measurements of TOA brings on gradually decreasing residuals, 
variances and covariances of all parameters. On the other hand, 
we show that any
low-frequency, correlated noise of terrestrial or/and astrophysical origin 
present causes the residuals, 
variances, and covariances of certain parameters to increase with time. 
Hence, the low frequency noise corrupts our observations and reduces
experimental possibilities for better tests of General Relativity Theory. At the
same time, the rate of growth of residuals and variances of parameters can 
give a valuable information about the red noise itself. We also 
treat in detail the
influence of a polynomial drift of noise on the residuals and fitting 
parameters in
order to avoid confusion with red noise without the polynomial drift. Results
of the analitic analysis are used for discussion of a statistic describing
stabilities of kinematic (PT) and dynamic (BPT) pulsar time scales.          
\end{abstract}

\baselineskip=20pt
\section{Introduction}

Timing observations of single and, especially, binary millisecond pulsars are 
widely
recognized as being extremely important for progressing a number of branches of modern
astronomy and time keeping metrology. In particular, implication of pulsar 
timing for testing General Relativity in the strong-field regime (Taylor \& 
Weisberg 1982, 1989), creation of a new astronomical time scale based on the
high-stable rotation of millisecond pulsars (Ilyasov {\it {\it et al.}} 1989,
Kaspi {\it {\it et al.}} 1994)as well as setting up the upper limit on the energy density of
stochastic gravitational wave background in early universe (McHugh {\it {\it et al.}}
1996, Kopeikin 1997a, and references therein) are among the most successful 
and stimulating recent achivements in this area of research.    

The accuracy of pulsar timing observations is now approaching $\sim$ 100
nanoseconds. Such high precision requires construction of adequate data
processing algorithms taking proper account of the relevant physical effects 
which can contribute to the timing signal being processed. Significant 
theoretical understanding of
relativistic celestial mechanics of binary systems has been made since the
discovery by Hulse and Taylor (1975) 
of the first binary pulsar PSR B1913+16 (Damour {\it {\it et al.}} 1989,
Damour \& Sch\"afer 1988, Ohta \& Kimura 1988, Sch\"afer \& Wex 1993, Kopeikin \&
Potapov 1994, Sch\"afer 1995, Wex 1995).
Presently, one is able to predict both secular evolution and periodic 
perturbations of the
binary system's orbit up to the $2\frac{1}{2}$ post-Newtonian 
approximation (PNA) where emission
of gravitational waves starts to influence the orbital dynamics 
(Grishchuk \& Kopeikin 1983, Damour 1983). Presently, efforts are being made
towards developing theory of motion of binary systems in 3 PNA and 
$3\frac{1}{2}$ PNA of General Relativity (Iyer \& Will 1995, Jaranowski 
\& Sch\"afer 1997, Jaranowski 1997, Rieth 1997).     

It is worth remembering that processing of pulsar timing data comprised of
topocentric time of arrivals (TOA) of pulsar pulses is based on the
$\chi^2$ minimization procedure of fitting observational parameters of 
the pulsar to
the adopted model of the observed TOA. This includes polar motion corrections
UT1-UTC given by the International Earth Rotation Service (IERS), the
general relativistic model of motion of the observer around the barycenter
of the Solar system (Standish 1990) and the pulsar around the
barycenter of the binary system (Damour {\it {\it et al.}} 1989), the
post-Newtonian transformations between different time scales (Brumberg
Kopeikin 1990, Fukushima 1995) used in the model, and the law of
propagation of electromagnetic signals in gravitational fields 
(Shapiro 1964), as well as the interstellar and interplanetary medium 
(Rickett 1990, 1996). The overall model is rather sophisticated and presently
exists in the form of two independently developed computer programs: TEMPO 
(Taylor \& Weisberg 1989), and TIMAPR (Doroshenko \& Kopeikin 1990, 1995), 
both being available on the World Wide Web. It is worthwhile pointing out
that TIMAPR is based
on the unique theoretical approach for construction of relativistic time scales
and reference frames in the Solar system developed in (Kopeikin 1988, 
Brumberg \& Kopeikin 1989, 1990). 
 
Usually, the procedure for estimating pulsar parameters is based on the
premise that white noise dominates in TOA residuals. However, long-term 
monitoring of certain pulsars has revealed the presence of non-white
component of noise of astrophysical origin as well (Cordes \& Downs 1985, 
Kaspi {\it {\it et al.}} 1994, D'Allesandro {\it {\it et al.}} 1996). It is customary
call
such correlated noise as a coloured Gaussian (or, simply, as "red") one because 
its spectrum diverges at zero
frequency ("the infra-red catastrophe"). At low frequencies the red noise has a
non-flat spectrum and can be described in framework of single- or multiple- 
component power-law model. The lower the timing activity of the pulsar, 
the further toward low frequencies one must look in order to detect
the red noise in the spectrum of its TOA residuals. Although being fairly
difficult for detection, red noise contains invaluable information about
diversity of physical processes which take place in the neutron star interior, 
the interstellar medium, early universe, and man-made terrestrial clocks
(Lorimer 1996). 
For this reason, developing a rigorous procedure for accounting for the
red noise component in TOA residuals is worthwhile. Physicists have been
working on the problem of adequate treatment of red noise for a long time (see,
for example, Stratonovich \& Sosulin 1966, Van Trees 1968, Hooge 1976, 
Planat {\it {\it et al.}} 1996). However, one of the first attempts in developing 
proper
statistical procedure of fitting pulsar parameters
in the presence of low-frequency, red noise has been undertaken only 
recently by
Bertotti {\it {\it et al.}} (1983), and Blandford {\it {\it et al.}} (1984) 
in order to obtain unbiased estimates of the pulsar's 
parameters as well as upper limit on the energy density of the GWB
radiation. There is a need to further
improve on probabilistic models of the red noise and 
methods of estimation of its characteristics containing
essential information about long-term processes having both geophysical and
astrophysical origins, including the stochastic background of primordial
gravitational radiation. The present state-of-the
art of statistical analysis of pulsar timing observational data in the presence
of red noise has not yet
reached the required level of clarity and completeness, and more elaborate
methodology has to be developed. This point has been especially stressed by
Bertotti {\it {\it et al.}} (1983), Hogan \& Rees (1984), Blandford {\it {\it et al.}}
(1984), and Kopeikin (1997a, 1997b). 

Systematic studies of timing noise
in post-fit residuals was started by Groth (1975) and Cordes (1978, 1980) 
(see also 
Cordes \& Helfand (1980), Cordes \& Greenstein (1981), and
Cordes \& Downs (1985)), who put forward a random walk noise model along
with the use of orthonormal polynomials for analyzing properties of the
noise in the time domain. These investigations
stimulated development of a somewhat equivalent approach for estimating the
spectral power of the low-frequency noise which was evaluated in the
Fourier frequency domain and non-uniformly sampled data by Deeter \& Boynton 
(1982) and Deeter (1984). The power spectrum technique was later applied 
to pulsar timing data
by Deeter {\it {\it et al.}} (1989). This formalism was worked further 
into a practical
procedure by Stinebring {\it {\it et al.}} (1989) and extensively
employed by Kaspi {\it {\it et al.}} (1994), and Thorsett \& Dewey 
(1996) for setting an upper limit on the cosmological
parameter $\Omega _{g}$ using the Neyman-Pearson test of statistical
hypothesis. McHugh {\it {\it et al.}} (1996) recently refined this procedure
to make it even more rigorous using the approach based on the Bayesian
statistic. 

It is worth noting that the analysis in the time domain is more informative
than that in the frequency domain. This is because noise
contains usually both stationary and non-stationary components, and spectral
analysis of the noise in frequency domain is adequate if and only if the
noise itself is stationary (or its increments). For this reason, the procedure 
advanced by Stinebring 
{\it {\it et al.}} (1989) is principally restricted by the implicit
assumption on stationarity of timing noise and, therefore, its implications
for processing real observational data are jeopardized. Fortunately, we have
been able to prove (Kopeikin 1997a) and reconfirm in the present paper 
that under rather general circumstances
the procedure of fitting of pulsar spin and orbital parameters acts as a
low-frequency filter of the non-stationary component of noise which means
that pulsar post-fit residuals bear, in fact, only the stationary component of 
noise and
the spectral analysis in the Fourier domain is legitimate. Nevertheless,
non-stationary part of noise intrudes into observed values of pulsar spin-down
parameters, increases their variances, and bringing on large correlations
with other parameters.

Another thing to notice is that inherent to Deeter-Boynton's procedure are
certain limitations, which must be clearly understood and accounted for in
practical implementations. The foremost restriction is that it provides a
satisfactory guide to the selection as well as construction of practical
estimators of a power spectrum of noise as long as the noise is
approximately represented by a {\rm (2r)-th} power law; that is, then it is
possible to write the noise spectrum down as $S(f)=K_{{\rm 2r}}f^{-{\rm 2r}},$
where ${\rm r}=1,2,3,...$, and any of $K_{{\rm 2r}}$ is a constant. Let us note
that the noise in question can be treated as a random walk in rotational phase,
frequency, or frequency derivative of the pulsar. Stationary
stochastic processes possessing power spectra with odd spectral indices $%
S(f)=K_{{\rm 2r+1}}f^{-{\rm 2r-1}},$ $({\rm r}=0,1,2,...)$ are called flicker
noise and require
development of a more advanced statistical approach. A possible way towards
the development of such an approach in the frequency domain was 
explored by Blandford {\it {\it et al.}} (1984). However, it
is not as far-going as one would desire. From the mathematical point 
of view the difficulty is due to the existence
of an algebraic singularity in the spectrum of low-frequency noise as
frequency approaches the point $f=0.$ When the spectral index of the noise
is large enough, the singular behavior of the power spectrum leads to 
formally divergent integrals describing timing residuals and variances of
measured parameters. To avoid this problem, a special regularization
procedure must be applied to these integrals to ascribe them definite
numerical values and, as a consequence, a real physical meaning. There are 
several 
known methods of dealing with regularization of functionals with
algebraic singularities. Blandford {\it {\it et al.}} (1984)
used the simplest procedure of regularization based on the effective
spectral cutoff of the divergent integrals. As a consequence, their
estimators of spectral power of the noise as well as variances of pulsar
parameters depend on the lower cut-off frequency, $f_{\min },$ and two
numbers, $A,$ and, $N^{*},$ which are used for minimizing root-mean-square
error in rotational frequency through the post-fit residuals. The cutoff
frequency $f_{\min }$ and numbers $A$ and $N^{*}$ are not constant and
depend on the total span of observations, making temporal behavior of the
post-fit residuals and variances of parameters rather uncertain. In
addition, the simple cutoff of divergent integrals leads to enormous leakage
of the spectral power from the low-frequency tails of the estimators and as
a consequence to the wrong determination of the magnitude $K_{{\rm m}}$ of
the noise power spectrum.

We have suggested tackling the problem of divergent integrals using the 
theory of
generalized functions (Gel'fand \& Shilov 1964) as the most powerful and
simple in dealing with functions having singular spectra. For instance, this
theory has been successfully applied in Quantum Renormalization Theory 
(Damour 1975) and for calculations of high-order relativistic equations of
motion of binary pulsars (Damour 1983a, Sch\"afer 1985). Quite
recently the theory of generalized functions has been applied by Kopeikin 
(1997a, 1997b) for
development of an adequate treatment of low-frequency timing noises with
negative integer spectral indices. In particular, we have discovered in these
papers for the first time that flicker noise caused by the stochastic background
of primordial gravitational radiation in early universe leads to the 
appearance of
specific logarithmic terms in the autocovariance functions as well as in
variances of spin parameters of observed pulsars. This result is 
re-confirmed and considerably extended in the present paper accounting for
flicker noise in rotational phase and frequency as well.

As a final critical remark, let us point out that the problem of proper
estimation of variances of orbital parameters of binary pulsars in the
presence of low-frequency noise has not been yet discussed in full detail. 
Therefore, 
the main goal of the present paper is to provide a self-consistent
theoretical and numerical study of this problem. Here we use the results
of our previous work (Kopeikin 1997a) in which the general statistical model of
red noise has beeen developed.  The plan for the present paper is as follows. 
In the next
section we describe an analytical timing model which is used in the
subsequent discussion. The procedure of estimation of variances of measured
parameters in the presence of low-frequency red noise is outlined in section
3. A brief description of model of a red noise and its autocovariance 
function are given in
section 4. Section 5 explains computational aspects of our algorithm.
Polynomial drift of timing noise is discussed in section 6. 
The drift-free noise model is employed in section 7 for analytical
evaluation of variances of spin and orbital parameters of a binary pulsar in
the presence of white and the low-frequency noises. Finally, the analytic
results are used for studying stability of kinematic (PT) and dynamic (BPT)
pulsar time scales in section 8. 

To complete the analysis the study of spectral 
sensitivity of binary pulsars in different frequency bands of the noise 
spectrum has to be done. This requires more work which is currently in progress. 
These results will be published elsewhere.

\section{Timing Model}

We consider a simplified, but still realistic model of arrival time
measurements of pulses from a pulsar in a binary system. It
is assumed that the orbit is circular, and
the pulsar rotates around its own axis with angular frequency $\nu _{p}$
which slows down due to the electromagnetic (or whatever) energy losses. It
is also taken into account that the orbital frequency of the binary system, $%
n_{b},$ and its projected semimajor axis, $x,$ have a
secular drift caused by emission of gravitational waves from the binary (Peters
\& Mathews 1963, Peters 1964) bringing about the gravitational radiation 
reaction force 
(Damour 1983a, Grishchuk \& Kopeikin 1983), radial acceleration 
(Damour \& Taylor 1991, Bell \& Bailes 1996), and proper motion of the
binary in the sky (Kopeikin 1996).

The moment, ${\cal T}$ , of emission of the ${\cal N}$-th pulsar's pulse
relates to the moment, $t,$ of its arrival, measured at the infinite
electromagnetic frequency, by the equations (Damour \& Taylor 1992, Kopeikin
1994): 
\begin{equation}
D\left[{\cal T}+x\sin \left( n_{b}{\cal T}+\sigma \right)\right]
 =t+\phi_0(t)+\phi_1(t),
\label{1.4}
\end{equation}
\begin{equation}
t=\tau^{\ast} +\Delta _{C}+\Delta _{R\odot }+\Delta _{\pi \odot }+\Delta _{E\odot
}+\Delta _{S\odot }.  \label{1.4a}
\end{equation}
We use the following notations:

\begin{itemize}
\item  ${\cal T}$ - pulsar time scale,

\item  $t$ - barycentric time at the barycenter of the Solar system,

\item  $\tau^{\ast} $ - topocentric time of observer,

\item  $\Delta _{C},$ $\Delta _{R\odot },$ $\Delta _{\pi \odot },$ $\Delta
_{E\odot },$ $\Delta _{S\odot }$ - clock and astrometric corrections (Taylor
\& Weisberg 1989, Doroshenko \& Kopeikin 1990, 1995) which one assumes to be
known precisely,

\item  $D$ - Doppler factor gradually changing due to the acceleration and
proper motion of the binary system in the sky \footnote
{$D=\frac{1+\frac{V_R}{c}}{\sqrt{\left(1-\frac{V^2}{c^2}\right)}}$, 
where $V_R$ and
$V$ are correspondingly the
relative radial and total velcities of the binary system barycenter with 
respect to the barycenter of the Solar system} 

\item  $\sigma $ - initial (constant) orbital phase,

\item  $n_{b}$ - orbital frequency $(n_{b}=2\pi /P_{b})$,

\item  $i$ - angle of inclination of the orbit to the line of sight,

\item  $x$ - projected semimajor axis $a_{p}$ of the pulsar's
orbit $(x=a_{p}\sin i/c)$,

\item  $c$ - speed of light,

\item  $\phi_0 (t)$ - the gaussian noise of TOA measuring errors,

\item  $\phi_1 (t)$ - low-frequency gaussian noise caused by the long-term 
instabilities of terrestrial clocks, effects in propagation of radio signals in
the interstellar medium, and stochastic background of primordial gravitational
waves.
\end{itemize}

We also suppose that observations start at the time $t_{0}=0.$ Changing of
the starting time of observations from $t_{0}=0$ to $t_{0}\neq 0$ is
achieved by the simple translation of time scale: $t\mapsto t-t_{0}.$ It is
worth emphasizing that for nearly circular orbits the initial orbital phase $%
\sigma =\omega _{0}-n_{b}{\cal T}_{0}$ where ${\cal T}_{0}$ is a fiducial
moment of time and $\omega _{0}$ is the longitude of periastron. This
relationship remains valid up to the first order of magnitude in
eccentricity and represents an excellent approximation in the case of binary
pulsars like PSR B1855+09 or PSR J1713+0747.

The rotational phase of the pulsar is given by the polynomial in time 
\begin{equation}
{\cal N}(t)=\nu _{p}{\cal T}+\frac{1}{2}\stackrel{.}{\nu }_{p}{\cal T}^{2}+%
\frac{1}{6}\stackrel{..}{\nu }_{p}{\cal T}^{3}+\frac{1}{24}\stackrel{...}{%
\nu }_{p}{\cal T}^{4}+\frac{1}{120}\stackrel{....}{\nu }_{p}{\cal T}%
^{5}+\nu_p\phi_2({\cal T})+O[{\cal T}^6],   \label{1.5}
\end{equation}
where $\nu _{p},$ $\stackrel{.}{\nu }_{p},\stackrel{..}{\nu }_{p},$ {\it etc.%
} are pulsar's rotational frequency and its time derivatives all referred to
the epoch ${\cal T}=0,$ the term $O[{\cal T}^6]$ denotes high order 
derivatives
of the rotational phase, and $\phi_2({\cal T})$ is the intrinsic pulsar timing 
noise in either
rotational phase, frequency, or frequency derivative. Solving iteratively 
equation (\ref{1.4}) with respect to $%
{\cal T}$ and substituting ${\cal T}$ for the right hand side of equation (%
\ref{1.5}) gives a relationship between two observable quantities ${\cal N}$
and $t$: \bigskip 
\begin{eqnarray}\label{1.7q}
{\cal N}(t)&=&{\cal N}_{0}+\nu t+{\frac{1}{2}}\stackrel{.}{\nu } t^2+
{\frac{1%
}{6}}\stackrel{..}{\nu }t^3+{\frac{1}{24}}\stackrel{...}{\nu }t^4
\\\nonumber\\\nonumber\mbox{}&& +{\frac{%
1}{120}}\stackrel{....}{\nu }t^5-  
\nu (x+\stackrel{.}{x}t+{\frac{1}{2}}\stackrel{..}{x}t^2+{\frac{1}{6}}%
\stackrel{...}{x}t^3)\sin (\sigma +n_{b}t+{\frac{1}{2}}\dot{n}_{b}t^2+ 
{%
\frac{1}{6}}\ddot{n}_{b}t^3)+\nu\;\epsilon (t)\;,
\bigskip   
\end{eqnarray}
with
\begin{eqnarray}
\epsilon(t) &\stackrel{def}{=}&\phi_0(t)+\phi_1(t)+\phi_2(t)\;,
\label{1.7aa}
\end{eqnarray}
where ${\cal N}_{0}$ is the initial rotational phase of the pulsar $({\cal N}%
_{0}\simeq -\nu t_{0})$; $\nu ,$ $\stackrel{.}{\nu },$ $\stackrel{..}{\nu }%
,...$ are the pulsar's rotational frequency and its time derivatives at the
initial epoch $t_{0};$ $x,$ $\stackrel{.}{x},$ $\stackrel{..}{x},...$ are
the projected semimajor axis of the orbit and its time derivatives at the
epoch $t_{0};$ $\sigma ,$ $n_{b},$ $\dot{n}_{b},$ $\ddot{n}_{b},...$ are the
pulsar's orbital initial phase, orbital frequency and its time derivatives
at the epoch $t_{0}.$ All non-linear terms of order $x^{2},x\epsilon ,$ $%
\epsilon ^{2},$ $\dot{\nu}x$, $\ddot{\nu}x$, etc. are negligible and 
have been omitted from (\ref{1.7q}).

Let us underline that secular variations (time derivatives) of pulsar
parameters include not only contributions caused by the dissipative physical
mechanisms like emission of gravitational waves by the binary pulsar,
but also depend on the kinematical effects caused by the radial
acceleration and proper motion of the pulsar in the sky (Damour \& Taylor
1991, Bell \& Bailes 1996, Kopeikin 1994, 1996). 

Another important point which is usually never stressed, but rather crucial, 
is the treatment of pulsar timing noise $\phi_2({\cal T})$. The
noise in question contributes to random jumps of the pulsar's phase at the
moment ${\cal T}$ of emission of pulse. However, observer at the Earth receives
the pulse, and consequently, information about the noise function $\phi_2$ much
later after the radio pulse has moved across the distance separating the pulsar
from observer. Thus, the observer is allowed to think about the pulsar timing
noise as starting long time ago and the question arises about how to account for
this nuisance "memory" effect. Actually, the "memory" effect problem is not a 
specific of the timing noise only. It exists in treatment of observations 
being corrupted by any 
low frequency noise. As far as we know the only paper, where the "memory" effect
in pulsar timing data processing has been discussed, is (Kopeikin 1997b).
Therein, we have proved that the "memory" effect is negligibly small under  
normal circumstances.
Namely for this reason we have deliberately omitted any explicit dependence on 
time ${\cal T}$
in formulae (\ref {1.7q}), (\ref {1.7aa}) where the only really important time 
argument is the Solar system barycentric time $t$.       

\section{Procedure of Estimation of Pulsar Parameters}

We assume that all observations of the binary pulsar are of a similar quality
and weight.
Then one defines the timing residuals $r(t)$ as a difference between the
observed number of the pulse, ${\cal N}^{obs},$ and the number ${\cal N}%
(t,\theta ),$ predicted on the ground of our best guess to the prior unknown
parameters of timing model (\ref{1.7q}), divided by the pulsar's rotational
frequency $\nu $, that is\bigskip 
\begin{equation}
r(t,\theta )=\frac{{\cal N}^{obs}-{\cal N}(t,\theta )}{\nu },\bigskip
\label{1.8a}
\end{equation}
where $\theta =\{\theta _{a},a=1,2,...k\}$ denotes a set of $k$ measured
parameters $[k=14$ in the model (\ref{1.7q})] which are shown in Table
{\ref{tab:par}}. It is worth noting that hereafter we use for the reason of 
convinuence the time argument $u=n_b t$.
\begin{table*}
\begin{minipage}{140mm}
\centering
\scriptsize\caption{List of the basic functions and parameters used in the fitting 
procedure. Spin parameters $\delta {\cal N}_{0}, \delta {\nu },
\delta\stackrel{.}{\nu }, \delta\stackrel{..}{\nu }, 
\delta\stackrel{...}{\nu }, \delta\stackrel{....}{\nu },$ fit rotational 
motion of the pulsar around its own axis. Keplerian parameters $\delta{x},
\delta\sigma, \delta{n}_{b}$ fit the Keplerian orbital motion of the pulsar 
about barycenter of the binary system. Post-Keplerian parameters 
$\delta\stackrel{.}{x}, \delta\stackrel{..}{x}, \delta\stackrel{...}{x}, 
\delta\stackrel{.}{n}_{b}, \delta\stackrel{..}{n}_{b}$ fit small observable
deviations of the pulsar's orbit from the Keplerian motion caused by the
effects of General Relativity, radial acceleartion, and proper motion
of barycenter of the binary system with respect to the observer}
\label{tab:par}
\vspace{5 mm}
\begin{tabular}{|ll@{\hspace{5 cm}}ll|}
\hline \\ \\
Parameter &&& Fitting Function \\ \hline
&  \\ 
$\beta _{1}={{\frac{\delta {\cal N}_{0}}{\nu }}}$&&& $\psi _{1}(t)=1$ \\ 
&  \\ 
$\beta _{2}={\frac{1}{n_{b}}}{{\frac{\delta \nu }{\nu }}}$ &&& $\psi
_{2}(t)=u$ \\ 
&  \\ 
$\beta _{3}={\frac{1}{2n_{b}^{2}}}{{\frac{\delta \stackrel{.}{\nu }}{\nu }}}%
 $ &&& $\psi _{3}(t)=u^{2}$ \\ 
&  \\ 
$\beta _{4}={\frac{1}{6n_{b}^{3}}}{{\frac{\delta \stackrel{..}{\nu }}{\nu }}}%
$ &&& $\psi _{4}(t)=u^{3}$ \\ 
&  \\ 
$\beta _{5}={\frac{1}{24n_{b}^{4}}}{{\frac{\delta \stackrel{...}{\nu }}{\nu }%
}}$ &&& $\psi _{5}(t)=u^{4}$ \\ 
&  \\ 
$\beta _{6}={\frac{1}{120n_{b}^{5}}}{{\frac{\delta \stackrel{....}{\nu }}{%
\nu }}}$ &&& $\psi _{6}(t)=u^{5}$ \\ 
&  \\ 
$\beta _{7}=-\delta x\sin \sigma -\delta \sigma x\cos \sigma $ &&& $\psi
_{7}(t)=\cos u$ \\ 
&  \\ 
$\beta _{8}=-\delta x\cos \sigma +\delta \sigma x\sin \sigma $ &&& $\psi
_{8}(t)=\sin u$ \\ 
&  \\ 
$\beta _{9}={\frac{1}{n_{b}}}\left( -\delta \stackrel{.}{x}\cos \sigma
+\delta n_{b}x\sin \sigma \right) $ &&& $\psi _{9}(t)=u\sin u$ \\ 
&  \\ 
$\beta _{10}={\frac{1}{n_{b}}}\left( -\delta \stackrel{.}{x}\sin \sigma
-\delta n_{b}x\cos \sigma \right) $ &&& $\psi _{10}(t)=u\cos u,$ \\ 
&  \\ 
$\beta _{11}={\frac{1}{2n_{b}^{2}}}\left( -\delta \stackrel{..}{x}\sin
\sigma -\delta \stackrel{.}{n}_{b}x\cos \sigma \right) $ &&& $\psi
_{11}(t)=u^{2}\cos u$ \\ 
&  \\ 
$\beta _{12}={\frac{1}{2n_{b}^{2}}}\left( -\delta \stackrel{..}{x}\cos
\sigma +\delta \stackrel{.}{n}_{b}x\sin \sigma \right) $ &&& $\psi
_{12}(t)=u^{2}\sin u$ \\ 
&  \\ 
$\beta _{13}={\frac{1}{6n_{b}^{3}}}\left( -\delta \stackrel{...}{x}\cos
\sigma +\delta \stackrel{..}{n}_{b}x\sin \sigma \right) $ &&& $\psi
_{13}(t)=u^{3}\sin u$ \\ 
&  \\ 
$\beta _{14}={\frac{1}{6n_{b}^{3}}}\left( -\delta \stackrel{...}{x}\sin
\sigma -\delta \stackrel{..}{n}_{b}x\cos \sigma \right) $ &&& $\psi
_{14}(t)=u^{3}\cos u$ \\ \\
\hline  
\end{tabular}
\end{minipage}
\end{table*}\\

If a numerical value of the parameter $\theta _{a}$ coincides with its
true physical value $\hat{\theta}_{a}$, then the set of residuals would
represent a physically meaningful noise $\epsilon (t)$, {\it i.e. }
\begin{equation}
r(t,\hat{\theta})=\epsilon (t).
\end{equation}
In practice, however, the true values of parameters
are not attainable and we deal actually with their least square estimates $%
\theta _{a}^{*}.$ Therefore, observed residuals are fitted to the expression
which is a linear function of corrections to the estimates $\theta _{a}^{*}$
of a priori unknown true values of parameters $\hat{\theta}_{a}$. From a
Taylor expansion of the timing model in equation (\ref{1.7q}), and the
fact that $r(t,\hat{\theta})=\epsilon (t)$ one obtains 
\begin{equation}
r(t,\theta ^{*})=\epsilon (t)-{\displaystyle \sum_{a=1}^{14}}\beta _{a}\psi _{a}(t,\theta
^{*})+O(\beta _{a}^{2}),  \label{1.9}
\end{equation}
where the quantities $\beta _{a}\equiv \delta \theta _{a}=\theta _{a}^{*}-%
\hat{\theta}_{a}$ are the corrections to the presently unknown true values of
parameters, and $\psi _{a}(t,\theta ^{*})=\left[ \frac{\partial {\cal N}}{%
\partial \theta _{a}}\right] _{\theta =\theta ^{*}}$ are basic fitting
functions of the timing model.

In the following it is more convenient to regard the increments $\beta _{a}$
as new parameters whose values are to be determined from the fitting
procedure. The parameters $\beta _{a}$ and fitting functions are summarized in
Table {\ref{tab:par}} with asterisks omitted and time $t$ is replaced for convenience by
the function $u=n_{b}t$ which is the current value of orbital phase. It is
worth emphasizing that the basic functions $\psi _{2a-1},$ ($a=1,...,7)$ are
even , and $\psi _{2a},$ $a=1,...,7$ are odd. We restrict the model to $14$
parameters since in practice only the first several parameters of the model
are significant in fitting to the rotational and orbital phases over the
available time span of observations. It is also important to understand that 
the smaller amount of fitting parameters one takes, the more significant 
the contribution of non-stationary part of low frequency noise (see discussion
at the end of this section).  

Now suppose that we measure $m$ equally spaced and comparably accurate
arrival times each orbit for a total of $N$ orbital revolutions, so we have $%
mN$ residuals $r_{i}\equiv r(t_{i}),$ $i=1,...,mN.$ Standard least squares
procedure (Bard 1974) gives the best fitting solution for estimates
of the parameters $\beta _{a}$ \bigskip 
\begin{equation}
\beta _{a}({\it T})={\displaystyle \sum_{b=1}^{14}}{\displaystyle 
\sum_{i=1}^{mN}}L_{ab}^{-1}\psi
_{b}(t_{i})\epsilon (t_{i}),\qquad a=1,...,14,\bigskip  \label{1.11}
\end{equation}
\bigskip where the matrix of information is \bigskip 
\begin{equation}
L_{ab}({\it T})={\displaystyle \sum_{i=1}^{mN}}\psi _{a}(t_{i})\psi _{b}(t_{i}),\bigskip
\label{1.11a}
\end{equation}
the matrix $L_{ab}^{-1}$ is its inverse, and ${\it T}=NP_{b}$ is a total
span of observational time. Matrices $L_{ab}$ and $L_{ab}^{-1}$ are given up to
numerical factor $\frac{m}{2\pi}$ in
Tables {\ref{tab:2}}$\div${\ref{tab:5}}. 

Let the angular brackets denote an ensemble average over many
realizations of the observational procedure. Hereafter, we assume that 
the ensemble average of the noise $\epsilon(t)$ is equal to
zero \footnote {see Section 6 where the influence of time polynomial drift 
of the ensemble average of noise is treated in more detail}. 
Hence, the mean value of any parameter $\beta_{a}$ is equal to zero as
well, {\it i.e.}
\begin{equation}
<\epsilon(t)> =0 \hspace{1 cm}\longrightarrow \hspace{1 cm} <\beta _{a}>=0.
\label{mean}
\end{equation}
The covariance matrix 
$M_{ab}\equiv $ $<\beta _{a}\beta _{b}>$ of the
parameter estimates is now given by the expression \bigskip 
\begin{equation}
M_{ab}({\it T})={\displaystyle \sum_{c=1}^{14}}{\displaystyle 
\sum_{d=1}^{14}}L_{ac}^{-1}L_{bd}^{-1}\left[
{\displaystyle \sum_{i=1}^{mN}}{\displaystyle \sum_{j=1}^{mN}}
\psi _{c}(t_{i})\psi
_{d}(t_{j})R(t_{i},t_{j})\right] ,\bigskip  
\label{1.12}
\end{equation}
where $R(t_{i},t_{j})=$ $<\epsilon (t_{i})\epsilon (t_{j})>$ is the
autocovariance function of the stochastic process $\epsilon (t)$.
The covariance matrix is
symmetric $\left( M_{ab}=M_{ba}\right) ,$ elements of its main diagonal give
variations (or dispersions) of measured parameters 
$\sigma_{\beta _{a}}\equiv M_{aa}$=$<\beta_{a}^{2}>$, and the off-diagonal 
terms represent the degree of statistic covariance (or correlation) between 
them. Covariance matrices are given explicitly in Section 6.

Subtraction of the adopted model from the observational data leads to the
residuals which are dominated by the random fluctuations only. An expression
for the mean-square residuals after subtracting the best-fitting solution
for the estimates (\ref{1.11}) is given by the formula \bigskip 
\begin{equation}
<r^{2}({\it T})>=\frac{1}{mN}{\displaystyle \sum_{i=1}^{mN}}%
{\displaystyle \sum_{j=1}^{mN}}F(t_{i},t_{j})R(t_{i},t_{j}),\bigskip  
\label{1.12a}
\end{equation}
where the function 
\begin{equation}
F(t_{i},t_{j})=\delta _{ij}-{\displaystyle \sum_{a=1}^{14}}
{\displaystyle \sum_{b=1}^{14}}L_{ab}^{-1}\psi
_{a}(t_{i})\psi _{b}(t_{j}),\bigskip  \label{1.12b}
\end{equation}
is called the filter function (Blandford {\it {\it et al.}} 1984). These expressions
merely demonstrate that the amount of the background noise is reduced by the
fit for the pulsar's spin and orbital parameters (Bertotti {\it {\it et al.}} 1983, 
Blandford {\it {\it et al.}} 1984, Bard 1974). Thus, the observed
magnitude of residuals is on the average smaller than that of the noise.
This is because we have chosen the estimates $\theta _{a}^{*}$ for
parameters so as to make the residuals as small as possible. Let us note that
the more fitting parameters one has in the timing model, the smaller is the mean
amplitude of residuals.  

Another remarkable feature of (\ref{1.12a}) is that if the 
autocovariance function $R(t_{i},t_{j})$
contains products of terms of the form $\psi _{c}(t_{i})\times f(t_{j}),$ where $%
f(t_{j})$ is an arbitrary smooth function, they must disappear from the post-fit
residuals  (Kopeikin 1997a). The proof of the statement is based on two exact
equalities: \bigskip 
\begin{equation}
{\displaystyle \sum_{i=1}^{mN}}{\displaystyle \sum_{j=1}^{mN}}\delta _{ij}\psi
_{c}(t_{i})f(t_{j})={\displaystyle \sum_{i=1}^{mN}}\psi _{c}(t_{i})f(t_{i}), 
\label{ad1}
\end{equation}
\bigskip and\bigskip 
\begin{equation}
{\displaystyle \sum_{i=1}^{mN}}%
{\displaystyle \sum_{j=1}^{mN}}%
{\displaystyle \sum_{a=1}^{14}}%
{\displaystyle \sum_{b=1}^{14}}L_{ab}^{-1}\psi _{a}(t_{i})\psi _{b}(t_{j})\psi
_{c}(t_{i})f(t_{j})= 
{\displaystyle \sum_{j=1}^{mN}}%
{\displaystyle \sum_{a=1}^{14}}%
{\displaystyle \sum_{b=1}^{14}}L_{ab}^{-1}L_{ac}\psi _{b}(t_{j})f(t_{j})= 
{\displaystyle \sum_{j=1}^{mN}}%
{\displaystyle \sum_{b=1}^{14}}\delta _{bc}\psi _{b}(t_{j})f(t_{j})= 
{\displaystyle \sum_{j=1}^{mN}}\psi _{c}(t_{j})f(t_{j}).
\label{ad2}
\end{equation}

We have set up the hypothesis (Kopeikin 1997a) that the non-stationary part 
of any low-frequency noise with rational power-law spectrum can be represented 
as a sum of terms being products of a polynomial of 
time $g(t)=a_0+a_1 t+a_2 t^2+...$ by a smooth function of time $f(t)$. This 
has been
proved for a fairly general event of low-frequency noise being generated by a
shot noise random process (Kopeikin 1997b) and seems to be true, if not for
all, at least for a considerable number of stochastic processes. Taking
large enough number
of fitting spin-down parameters and corresponding fitting functions $\psi(t)$
one can represent
the non-stationary component of noise 
as a sum of terms of the form $\psi(t_i)\times f(t_j)$. Hence, residuals for the
example given do not contain the non-stationary component. In real practice,
observers fit usually for the first three spin-down parameters - initial
rotational phase, frequency, and frequency derivative. Such a procedure
eliminates from the
residuals any non-stationary noise components having spectral index $n \leq
6$. 

Moreover, we emphasize that the property of the fitting procedure
to filter out terms of the form of $\psi _{c}(t_{i})f(t_{j})$ does not
actually depend on whether observations are equally spaced. Hence, the
conclusion is that, if the autocovariance function has a non-stationary
component comprised of a sum of terms of the form $\psi _{c}(t_{i})\times 
f(t_{j})$,
one can always make the post-fit residuals depending only on the 
stationary part of the noise 
\begin{equation}
<r^{2}({\it T})>=\frac{1}{mN}%
{\displaystyle \sum_{i=1}^{mN}}%
{\displaystyle \sum_{j=1}^{mN}}F(t_{i},t_{j})R^{-}(t_{i},t_{j})=
-\frac{1}{mN}{\displaystyle \sum_{a=1}^{14}}{\displaystyle 
\sum_{b=1}^{14}}L_{ab}^{-1}\left[
{\displaystyle \sum_{i=1}^{mN}}{\displaystyle \sum_{j=1}^{mN}}
\psi _{a}(t_{i})\psi
_{b}(t_{j})R^{-}(t_{i},t_{j})\right] 
. \label{vlbi}
\end{equation}
For this reason, methods of spectral analysis in frequency domain can be 
applied without any restriction.

To find the asymptotic behavior of the residuals and elements of the covariance
matrix as functions of time (or a number of orbital revolutions $N$) one needs 
to restrain a model of the stochastic noise process.

\section{Model of Noise and Its Autocovariance Function}

We have already assumed that noise consists of algebraic sum of mutually 
uncorrelated
components of white and low-frequency noises (see Eq.({\ref{1.7aa})). White 
noise is due to measuring
uncertainty of TOA while the low-frequency noise has terrestrial and/or 
astrophysical origin and becomes significant only on long enough span of 
observational time. Investigation of its nature is one of the most important
problems of pulsar timing since it gives us a key to much better understanding
of the very deep fundamentals of physical laws governing the evolution of 
Nature on long time intervals. Especially important in this respect are 
observations of millisecond pulsars in binary systems having the best
attainable accuracy and, consequently, the low level of white noise component
in TOA residuals. 

White noise is stationary and has a flat constant spectrum
\begin{equation}
S(f)=h_0, \label{4.1}
\end{equation}
with $h_0$ being a fixed parameter determined from the measured level of white
noise on short time scales (see Eq. ({\ref{wpn1}}). The autocovariance 
function of white noise is 
\begin{equation}
R(\tau)=h_0 \delta(\tau), \label{4.2}
\end{equation}
where $\delta(\tau)$ is the Dirac delta function and $\tau = t_i - t_j$ with
$t_i, t_j$ being moments of TOA of the $i$-th and $j$-th observations.
  
The generalized model of low frequency noise was developed by 
Kopeikin (1997b) using the shot-noise approximation. It can
describe both random walk and flicker noises having, as a rule, 
different physical origins.
The autocovariance function of the noise includes both stationary and
non-stationary components which are treated on equal footing. 
We suppose that the autocovariance function of any stochastic
process under consideration is a function of real variable and can be split 
algebraically into stationary and non-stationary components \bigskip 
\begin{equation}
R(t_{i},t_{j})=R^{+}(t_{i},t_{j})+R^{-}(t_{i},t_{j}).\bigskip  \label{ogloed}
\end{equation}
The function $R^{+}(t_{i},t_{j})$ describes the non-stationary part of the
noise and $R^{-}(t_{i},t_{j})\equiv R^{-}(\left| t_{i}-t_{j}\right| )$ is
its stationary counterpart. The latter can be displayed in the
frequency domain through the cosine Fourier transform\bigskip 
\begin{equation}
R^{-}(\tau )=2\int_{0}^{\infty }S(f)\cos (2\pi f\tau )df,\hspace{20 mm}
\tau\equiv t_{i}-t_{j}\bigskip
\label{lapa}\\
\end{equation}
where $S(f)$ is the spectrum of $R^{-}(\tau )\bigskip $%
\begin{equation}
S(f)=2\int_{0}^{\infty }R^{-}(\tau )\cos (2\pi f\tau )d\tau .\bigskip
\label{papa}
\end{equation}
The stationary part of the red noise has a power-law spectrum, $S(f)$, 
being proportional to
$f^{-n}$ where $n$ is called the spectral index 
of the noise and in the present paper $n=1, 2, 3, 4, 5, 6$.
Corresponding autocovariance functions of the stationary components are 
represented by either polynomial of time (for random walk noise) or polynomial 
of time plus logarithmic function of time multiplied by another polynomial of 
time (for flicker noise). Nomenclature of the corresponding names of noises, 
their spectra, and stationary part of the autocovariance functions are given 
in Table {\ref{tab:1}} which is an extract from (Kopeikin 1997b).
Non-stationary parts of the autocovariance functions are expressed as a sum of 
terms of the form: ${\rm [basic}$ ${\rm %
function}$ ${\rm of}$ ${\rm fitting}$ ${\rm procedure}$ $\psi _{c}(t_{i}]%
{\rm )\times [a\ smooth\ function}$ $f(t_{j})]$ and are given below in Section
7. As mentioned above, such terms do not contribute to post-fit 
timing residuals. However, they do contribute to the fitted values of the 
pulsar's spin parameters. 
\begin{table*}
\centering
 \begin{minipage}{140mm}
 
  \scriptsize\caption{The spectra of timing noise and the stationary part of
  the autocovariance functions $R^{-}(t_i,t_j)$ used in the paper. Constant
   parameters $h_n$, where $n=1,2,...,6$, characterize the
   magnitude of the noise spectrum. The quantity $\tau=t_i-t_j$, and one
   introduces notation
   $\tau_{-}=-\tau$, if $\tau<0$, and $\tau_{-}=0$, if $\tau \geq 0$.}
   \label{tab:1}
   \centering
  \begin{tabular}{|llc|}
  \hline\\ \\
   Noise& Spectrum &Autocovariance Function\\ \\
   \hline \\   
 White Noise (WN)                            & $h_0$&$h_0 \delta(\tau)$\\ \\
 Flicker Noise in Phase (FPN)         & $h_1
                                    f^{-1}$&$-\frac{h_1}{\pi}\ln|\tau|$\\ \\
 Random Walk in Phase (RWPN)                 & $h_2f^{-2}$&$-h_2\tau_{-}$\\ \\
 Flicker Noise in Frequency (FFN)&$h_3
                              f^{-3}$&$\frac{1}{2\pi}h_3\tau^2\ln|\tau|$\\ \\
 Random Walk in Frequency (RWFN)             & $h_4
                                      f^{-4}$&$\frac{1}{6}h_4\tau_{-}^3$\\ \\
 Flicker Noise in Frequency Derivative (FSN) & $h_5
                            f^{-5}$&$-\frac{1}{24\pi}h_5\tau^4\ln|\tau|$\\ \\
 Random Walk in Frequency Derivative (RWSN)  & $h_6
                      f^{-6}$&$-\frac{1}{120}h_6\tau_{-}^5$\\ \\
 \hline
\end{tabular}
\end{minipage}
\end{table*}\\

Before going further, let us discuss in more detail computational aspects of 
our analytical approach for calculation of residuals and covariance matrices 
under the influence of different noise processes.  

\section{Computational Aspects}

In order to accomplish all calculations analytically we assume that the
number of observations is so large that any sum over observing points can be
approximated by the integral over the observing period ${\it T^{\ast}}$. 
For calculational convenience, 
we assume that observations are commenced at
the moment $t_0=-{\it T^{\ast}}/2$ and finished at $t_0={\it T^{\ast}}/2$ 
so that the interval of integration time is symmetric with respect to the 
origin of time scale. Such a shift of the time scale is always possible
and preserves the invariance of timing formula ({\ref{1.7q}}).
Moreover, we assume that observations are equally spaced with small enough 
interval
of time between successive observations $\triangle t=P_b/m$ where $m$
is the number of observations per one orbital period. A total 
number of orbital revolutions $N$ is supposed to be large ($N \geq 30$). 

Under the given conditions one may apply to any smooth function 
$f(t)\in C^\infty [-t_0,t_0]$ the
Euler-Maclaurin summation formula (2.9.15) given in textbook of 
Davis \& Rabinowitz (1984) and having the form of an asymptotic expansion
\begin{equation} 
{\displaystyle \sum_{i=1}^{mN}}f(t_i)=
{\displaystyle \sum_{i=1}^{mN}}f\left[ -t_0+(i-1)\triangle t\right] = 
\frac m{2\pi }{\displaystyle\int_{-\rm T}^{\rm T}}f(u)du+E(m,N),
\label{1.44} 
\end{equation}
where ${\rm T}\equiv{\pi N}$, and the equalities $n_b\triangle t/P_b=n_b du/(2\pi)$ 
and ${\rm T}=n_bt_0=n_b {\it T^\ast}/2$ have been used\footnote{New number
${\rm T}$ counts the span of observational time in terms of 
orbital revolutions of the
binary pulsar in question. When discussing observations of a single pulsar 
we treat the number $N$ as being counted in years}. Herein $E(m,N)$ is the 
error of approximation of the summation formula ({\ref{1.44}})
\begin{equation} 
E(m,N)=\frac 12\left[ f(-{\rm T})-f({\rm T})\right]   
+{\displaystyle \sum_{k=1}^{\infty}} 
\frac{B_{2k}}{(2k)!}\left( 
\frac{2\pi }{m}\right) ^{2k-1}\left[ f^{(2k-1)}({\rm T})-f^{(2k-1)}({-\rm
T})\right] , 
\label{1.444}
\end{equation}
where derivatives $f^{(k)}$ of the function $f(u)$ are taken with respect to 
the argument $u,$ and the constants $B_{2k}$ are called the 
{\it Bernoulli numbers} with the numerical values 
\begin{equation}
B_2=\frac 16,\qquad B_4=-\frac 1{30},\qquad B_6=\frac
1{42},\qquad B_8=-\frac 1{30},\hspace{10 mm}{\it etc.}
\label{1.445}
\end{equation}
It is not difficult to check that the error $E(m,N)$ is decreasing, at least as 
$(mN)^{-1}$ comparativly to the main term, or even faster.
It may be made small enough (less than 5
observations per orbital period
$m\geq 10$ and the number of orbital revolutions $N\geq 30$) (Wex 1997). 
We assume, in what follows, that this condition is
fulfilled and the residual terms $E(m,N)$ are negligibly small.

Because the autocovariance function consists of a linear combination of two terms
representing non-stationary and stationary components, the covariance matrix 
(\ref{1.12}) is split into two liearly independent terms as well:  
\begin{equation}
M_{ab}=M_{ab}^{+}+M_{ab}^{-}, \label{1.44a}
\end{equation}
where $M_{ab}^{+}$ depends only on $R^{+}(t_i,t_j)$ and $M_{ab}^{-}$
depends only on $R^{-}(t_i,t_j)$. Let us underline that the shift in the origin
of initial epoch of observations leads to the shift of arguments in the
autocovariance functions $t_i\rightarrow t_i+{\it T^\ast}/2,$ 
$t_j\rightarrow t_j+{\it T^\ast}/2.$
It leaves the stationary part of the autocovariance function 
$R^{-}(t_i,t_j)$ invariant, depending on the difference $t_i-t_j$ only, though
produces 
changes in the non-stationary part of $R^{+}(t_i,t_j)$ (see section 7 
for more detail).

Calculations of integrals depending on the non-stationary part $%
R^{+}(t_i,t_j)$ of the autocovariance function are relatively easy to handle. 
Let
us remind that the function $R^{+}(t_i,t_j)$ for any noise model is composed
of the finite sum of products of basic functions $\psi _l(t_i)\psi
_k(t_j),$ $(l,k\leq 6)$ in the event of random walk noise to which functions 
having the structure $\psi _l(t_i)\ln
(n_bt_j+{\rm T}),$ $l\leq 3$ are added in the event of flicker noise 
(Kopeikin 1997b). On one hand, for random walk the corresponding part of
the covariance matrix will be 
\begin{equation}
\begin{array}{r}
M_{ab}^{+}\sim 
{\displaystyle \sum_{c=1}^{14}}{\displaystyle \sum_{d=1}^{14}}%
L_{ac}^{-1}L_{bd}^{-1}\left[{\displaystyle \sum_{i=1}^{mN}}\psi
_c(t_i)\psi_l(t_i)\right] \left[ {\displaystyle \sum_{j=1}^{mN}}\psi
_d(t_j)\psi_k(t_j)\right]\\ \\  
=\left[ {\displaystyle \sum_{c=1}^{14}}L_{ac}^{-1}L_{cl}\right]
\left[ {\displaystyle \sum_{d=1}^{14}}L_{bd}^{-1}L_{dk}\right]
=\delta_{al}\delta_{bk},\qquad (l,k\leq 6),
\end{array} 
\label{1.45}
\end{equation}
where $\delta _{ab}$ is the unit matrix. On the other hand, 
in the event of flicker noise one has additional
contributions
\begin{equation}
\begin{array}{r}
M_{ab}^{+}\sim 
{\displaystyle \sum_{c=1}^{14}}{\displaystyle \sum_{d=1}^{14}}%
L_{ac}^{-1}L_{bd}^{-1}\left[ {\displaystyle \sum_{i=1}^{mN}}\psi
_c(t_i)\psi _l(t_i)\right] \left[{\displaystyle \sum _{j=1}^{mN}}\psi
_d(t_j)\ln (n_bt_j+{\rm T})\right]=\\ \\ 
\delta _{al}\left[ \frac {m}{2\pi }{\displaystyle \sum_{d=1}^{14 }}%
L_{bd}^{-1}{\displaystyle\int_{-{\rm T} }^{{\rm T}}}
\psi _d(u)\ln (u+{\rm T})du\right] ,\hspace{1 cm}(l\leq 3). 
\end{array}
\label{1.46}
\end{equation}

Relationships (\ref{1.45})-(\ref{1.46}) lead to the important
conclusion that, for the timing model under consideration, the non-stationary 
part of any autocovariance function
contributes only to the elements of the covariance matrix 
$M_{ab}$ with indices $a,b\leq 6$ for the random
walk noise, and to the elements of $M_{ab}=M_{ba}$ with $a\leq 3, b=1,2,...14$ 
for the flicker noise.
This specific behaviour of non-stationary part of autocovariance functions 
was noted partially by Groth (1975) in the context of his
investigation of random walk processes in pulsar timing based upon the usage
of orthogonal polynomials. Our derivation of the given result is, in fact,
more general as it does not depend on the particular choice of special functions
used for expansion of random process and is applicable both to random walk and 
flicker noises as well.

Calculations of integrals depending only on the stationary part of
autocovariance function are fulfilled after making a preliminary
transformation of independent variables (see Fig. {\ref{fig:1}}).
\begin{figure*}
\centerline{\psfig{figure=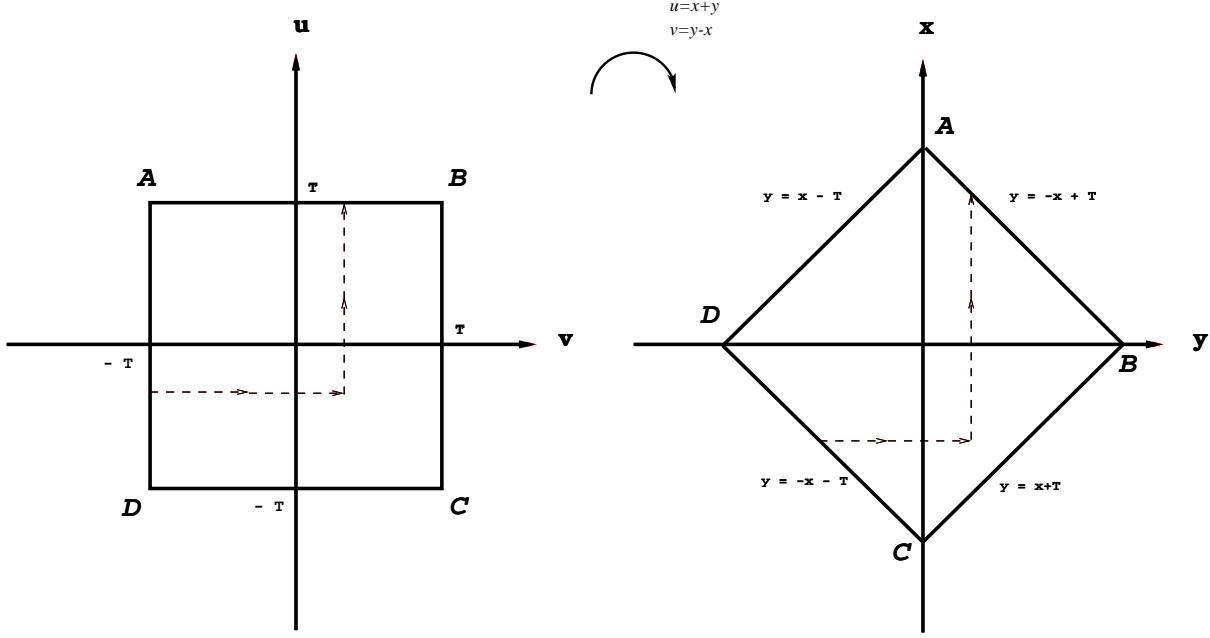,angle=270,width=16cm}}
\caption{Geometrical illustration showing change of independent 
          variables and transformation of the domain of integration in the
          phase space of
          two time arguments. Dashed arrowed lines indicate paths of
          integration.}
\label{fig:1}
\end{figure*}
Let us denote $u=n_b t_i,$ $v=n_b t_j,$ $t_i\neq t_j$ and make a transformation 
\begin{equation} 
x=\frac{u-v}{2}, \hspace{1cm} y=\frac{u+v}{2} .
\label{1.47}
\end{equation}
It is obvious that the stationary part $R^{-}(t_i,t_j)$ of the autocovariance 
function depends only on 
$|x|$. Hence, the part of the covariance matrix $M_{ab}^{-}$ is
transformed after using formula (\ref{1.47}) to the integral 
\begin{equation}
M_{ab}^{-}= 
{\displaystyle\int_{-{\rm T}}^{{\rm T}}}
{\displaystyle\int_{-{\rm T}}^{{\rm T}}}\psi _a(u)\psi _b(v)R^{-}(u-v)dudv=  
{\displaystyle \int_{0 }^{{\rm T}}}%
dx R^{-}(x){\displaystyle \int_{0 }^{x-{\rm T}}}dy A_{ab}(x,y), 
\label{1.48}
\end{equation}
where the symmetric matrix 
\begin{equation}
A_{ab}(x,y)\equiv -2\left[ 1+(-1)^{a+b}\right][\psi _a(u)\psi _b(v)+\psi
_a(v)\psi _b(u)].
\label{1.48aa}
\end{equation}
The result shows that in the analytical approximation under
consideration the elements of matrix $M_{ab}^{-}\neq 0$, if and only if
the basic functions $\psi _a(u)$ and $\psi _b(v)$ are both either odd or
even. In other words, any fitting parameter having odd number has no 
correlation with that having even number. 

For this reason, when calculating integrals by formula (\ref{1.48aa}) it is more 
convenient to re-group fitting parameters 
in order to represent the matrix $A_{ab}(x,y)$ into a block-diagonal 
form 
\begin{equation}
A_{ab}\hspace{1 cm}\Rightarrow \hspace{1 cm}\left( 
\begin{array}{cc}
A_1 & 0 \\ 
0 & A_2 
\end{array}
\right) , \label{1.48a}
\end{equation}
where the left upper block $A_1=A_{2a-1,2b-1}$ depends only on odd fitting 
functions, and the right lower block  
$A_2=A_{2a,2b}$ depends only on even ones $(a,b=1,...,7.$ Due to this
re-arrangement of parameters, the matrix $%
L_{ab}$ and consequently the inverse matrix $L_{ab}^{-1}$ are also reduced
to the block-diagonal form having the same structure as in (\ref {1.48a}).
The advantage of this re-arrangement is that one can work with each of the block
matrices independently, simplifying the calculations considerably. Nevertheless,
it is worthwhile to remind that such simplification can be achieved in the main
approximation only. If one took terms of higher order in the expansion
(\ref{1.44}) the re-arrangement of elements of matrix $M_{ab}^{-}$ would be 
worthless since odd and even fitting parameters get correlated. In such event 
one can make a progress in calculations only if numerical methods are implied.
This is the case one meets in a real practice.  

In the analytical approximation under consideration the integral (\ref{1.48}) 
was calulated using the enhanced version 2.2.3 of MATHEMATICA for 
Microsoft Windows. It turns out that the outcome
of calculation can be always expressed either as a polynomial and
trigonometric functions of time in the event of random walk, or as a polynomial,
trigenometric functions, and sine/cosine-integrals in the event of flicker
noise. Since the number of observations $mN$ has been assumed to be very
large all functions appearing in elements of covariance matrix (\ref{1.48}) 
have been expanded
into asymptotic series with respect to the small parameter 
$\varepsilon =\frac {1}{\pi%
N}\equiv \frac{1}{\rm T}.$ Then, only the first term of the expansions has been 
retained and all
residual terms have been abandoned. An explicit result of such calcualtion is,
for instance, the 
matrix of
information $L_{ab}=\frac{m}{2 \pi} C_{ab}$ given in Tables {\ref{tab:2}} 
and {\ref{tab:3}}. 
Analytical expressions for
residuals and elements of covariance matrices of 
pulsar's fitting parameters in presence of white
and red noises are given in Section 7.
 
\section{Treatment of Polynomial Drift of Timing Noise}

For mathematical convenience, in this section we shall assume that the number 
of spin-down parameters of timing model is equal to $M$ 
and the total number of fitting parameters is $K$ 
\footnote {In particular, the numbers $M$ and $K$ are equal to $M=6$ and 
$K=14$
respectively in the timing model described by equation (\ref {1.7q})}.
Up to now, we have been concerned mainly with the mathematical formulation of
procedure of estimation of pulsar's parameters under the assumption that noise
has only random fluctuations, that is equation (\ref {mean}) is true. 
However,
nonrandom variations (drifts) of the noise do exist and can be sometimes modeled
by deterministic polynomial functions of time as it takes place in description
of electromagnetic breaking of pulsar's rotation. Hence, it is conceivable
to imagine that the ensemble average of noise is modeled by a $(P-1)$-th degree
polynomial
\begin{equation}
<\epsilon (t)>={\displaystyle \sum_{k=1}^{P}}\alpha_k t^{k-1}=
{\displaystyle \sum_{k=1}^{M}}n_b^{-k+1}\alpha_k \psi_k(t)+
\left(\alpha_{M+1} t_i^M+...+\alpha_P t_i^{P-1} \right),
\label{aver}
\end{equation}
where $\alpha_k$ are constants which numerical values get smaller as $k$ 
increases.
From the definition of fitting parameters and fitting
functions as well as equation
(\ref {1.11}) it is easy to verify that the mean value of the parameters 
reads as
\begin{equation}
\begin{array}{ll}
<\beta _{a}({\it T})>&={\displaystyle \sum_{b=1}^{K}}%
{\displaystyle \sum_{k=1}^{P}}
{\displaystyle \sum_{i=1}^{mN}}L_{ab}^{-1}\alpha_k\psi_{b}(t_{i})%
t_i^{k-1}=
{\displaystyle \sum_{b=1}^{K}}%
{\displaystyle \sum_{k=1}^{M}}
{\displaystyle \sum_{i=1}^{mN}}L_{ab}^{-1}n_b^{-k+1}\alpha_k\psi_{b}(t_{i})%
\psi_{k}(t_{i})+
{\displaystyle \sum_{b=1}^{K}}%
{\displaystyle \sum_{k=M+1}^{P}}
{\displaystyle \sum_{i=1}^{mN}}L_{ab}^{-1}\alpha_k\psi_{b}(t_{i})%
t_i^{k-1}\\ \\&=
{\displaystyle \sum_{k=1}^{M}}n_b^{-k+1}\delta_{ak}\alpha_k+
{\displaystyle \sum_{b=1}^{K}}%
{\displaystyle \sum_{i=1}^{mN}}L_{ab}^{-1}\psi_{b}(t_{i})\left(%
\alpha_{M+1} t_i^M+...+\alpha_P t_i^{P-1} \right).
\label{aver1}
\end{array}
\end{equation}

Thus, we conclude that until number $P$ of the polynomial's coefficients 
(\ref {aver}) is less or equal than
number $M$ of spin-down parameters being used in deterministic timing model
it
changes only the mean values of the first $M$ parameters. In such event one can
make several more iterations in the procedure of minimization of square of 
residuals in
order to eliminate completely the polynomial drift from the ensemble 
average of the noise by means of re-shifting first $P$ spin-down parameters.
As a result, one will get $<r(t)>=0$. 

If $P > M$, then 
\begin{equation}
<r(t)>=\alpha_{M+1} t^M+...+\alpha_P t^{P-1} -%
{\displaystyle \sum_{a=1}^{K}}   %
{\displaystyle \sum_{b=1}^{K}}L_{ab}^{-1}\psi_a(t)\left[%
{\displaystyle \sum_{i=1}^{mN}}\psi_{b}(t_{i})(\alpha_{M+1} t_i^M+...+
\alpha_P t_i^{P-1}) %
\right].
\label{aver2}
\end{equation}
One recognizes that in this event $<r(t)>\neq 0$ though in practice 
it can be already within the limit of rounding error of numerical computations.
It is worth emphasizing that even if it is the case, some fitted parameters
$\beta_a$ can, nevertheless, have the mean value lying above the limit of the 
rounding error.
We conclude that it is impossible to make the mean value for the residuals 
and all
fitting parameters equals to zero if the number of spin-down parameters in
timing model
is less than number of coefficients in polynomial drift of the
ensemble average of noise. This conclusion holds true for both white and red
noise regimes.

Let us turn attention now to the calculation of residuals in the presence of
polynomial drift of the noise's ensemble average (\ref {aver}). The 
autocovariance function of noise reads now as
\begin{equation}
<\epsilon(t_i)\epsilon(t_j)>=R(t_i,t_j)+%
{\displaystyle \sum_{k=1}^{P}}%
{\displaystyle \sum_{l=1}^{P}} \alpha_k \alpha_l t_i^{k-1}t_j^{l-1},
\label{aver3}
\end{equation}
where $R(t_i,t_j)$ is the autocovariance function of the noise with zero-valued
ensemble average. Influnce of the function $R(t_i,t_j)$ on 
residuals and 
covariance matrix of fitting parameters is explicitly demonstrated in the next 
section. Calculation of ensemble average of residuals with making use of 
expression (\ref {aver3}) and the property that fitting procedure filters out all
terms of the form $\psi_a(t_i) f(t_j)$ yields
\begin{equation}
<r^{2}({\it T})>=%
\frac{1}{mN}%
{\displaystyle \sum_{i=1}^{mN}}%
{\displaystyle \sum_{j=1}^{mN}}F(t_{i},t_{j})%
\left[R^{-}(t_{i},t_{j})+{\displaystyle \sum_{k=M+1}^{P}}%
{\displaystyle \sum_{l=M+1}^{P}} \alpha_k \alpha_l t_i^{k-1}t_j^{l-1}\right].
\label{aver4}
\end{equation}
Thus, one can see that if the condition $P<M+1$ holds, then the polynomial drift
of noise does not contribute to the ensemble mean value of square of timing
residuals. In the opposite situation, when $P \geq M+1$, there extra
contribution exist and there is a danger of confusion of white noise having a
polynomial drift with a red noise without such a drift.

Regarding covariance matrix of parameters in presence of polynomial drfit of
noise it is not so difficult to generalization (\ref {1.12}) using 
expression (\ref {aver3}) for the autocovariance function of noise. It results in 
\begin{equation}
M_{ab}= M_{ab}^{-}+M_{ab}^{+}+\Delta M_{ab},
\label{aver5}
\end{equation}
where the stationary part, $M_{ab}^{-}$, the non-stationary part, $M_{ab}^{+}$, 
and the increment $\Delta M_{ab}$ of covariance matrix read, correspondingly, as
\begin{equation}
M_{pq}^{-}({\it T})={\displaystyle \sum_{c=1}^{K}}{\displaystyle 
\sum_{d=1}^{K}}L_{pc}^{-1}L_{qd}^{-1}\left[
{\displaystyle \sum_{i=1}^{mN}}{\displaystyle \sum_{j=1}^{mN}}
\psi _{c}(t_{i})\psi
_{d}(t_{j})R^{-}(t_{i},t_{j})\right] ,
\label{aver6}
\end{equation}
\begin{equation}
\begin{array}{ll}
M_{pq}^{+}({\it T})&={\displaystyle \sum_{c=1}^{K}}{\displaystyle 
\sum_{d=1}^{K}}L_{pc}^{-1}L_{qd}^{-1}
{\displaystyle \sum_{i=1}^{mN}}{\displaystyle \sum_{j=1}^{mN}}
\psi _{c}(t_{i})\psi
_{d}(t_{j}) R^{+}(t_{i},t_{j})+
{\displaystyle \sum_{k=1}^{M}}{\displaystyle \sum_{l=1}^{M}}%
n_b^{-k-l+2}\delta_{pk}\delta_{ql}\alpha_k \alpha_l\\ \\\mbox{}
&+2\left[{\displaystyle \sum_{k=1}^{M}}n_b^{-k+1}\delta_{pk}\alpha_k \right]
\left[{\displaystyle \sum_{c=1}^{K}}
{\displaystyle \sum_{j=1}^{mN}}L_{qd}^{-1}\psi _{d}(t_{j})
{\displaystyle \sum_{l=M+1}^{P}}\alpha_l t_j^{l-1}\right] ,
\end{array}
\label{aver7}
\end{equation}
\begin{equation}
\Delta M_{ab}({\it T})={\displaystyle \sum_{c=1}^{K}}{\displaystyle 
\sum_{d=1}^{K}}L_{ac}^{-1}L_{bd}^{-1}\left[
{\displaystyle \sum_{i=1}^{mN}}{\displaystyle \sum_{j=1}^{mN}}
 \psi _{c}(t_i)\psi_{d}(t_j)
{\displaystyle \sum_{k=M+1}^{P}}{\displaystyle \sum_{l=M+1}^{P}} 
 \alpha_k \alpha_l t_i^{k-1} t_j^{l-1}\right].
\label{aver8}
\end{equation}
Herein, we have assumed that the non-stationary part of autocovariance function
of noise $R^{+}(t_{i},t_{j})$ consists only of products $\psi_a(t_i) 
f(t_j)$ with $a \leq M$.

Analayzing the structure of relationships (\ref {aver5})-(\ref {aver8}) we
draw the following conclusions: 1) if the number of coefficients $P$ in the 
polynomial drift of noise is less than or equal to the number $M$ of spin-down 
parameters in the timing model, then it changes only non-stationary part 
$M_{ab}^{+}$ of the covariance matrix for
spin-down parameters; -2) if the number of coefficients $P$ in the 
polynomial drift of noise is more than or equal to the number $M$ of spin-down 
parameters in the timing model, then it influences the non-stationary part 
of the covariance matrix for 
spin-down and orbital parameters and gives rise to $\Delta M_{ab}$. 
Hence, if we do not account for the
polynomial drift properly it may worsen numerical boundaries for true 
estimates of variances and covariances of fitting parameters. 

At this point, it is worthwhile to emphasize that in order to make effects of polynomial
drift of noise as small as possible one has to introduce to the timing model as
many spin-down parameters as it is required by observational accuracy and
optimization of the least-square minimization procedure. 
     
\section{Residuals, Covariance Matrix, and Variances of 
   Measured Parameters}
\subsection{General Comments}

In what follows we assume that effect of fitting spin-down
parameters is so suppressive that the polynomial drift of the noise is
completely 
eliminated
from all residuals and mean values of parameters. When calculating
covariance
matrices of parameters, it is worth keeping in mind that if one takes 
only stationary
components of low frequency noise it is not sufficient to  demand the
variances of
all fitting parameters to be positive since the first several spin-down
parameters
can have variances with negative values (see Tables (9)-(16) for more
detail). 
Negative variances are, of course, physically meaningless. Hence, we
conclude
that it is incorrect to disregard the non-stationary part of red noise
for it
hampers the physical interpretation of these variances. It turns out
that, in order   
to obtain positive values for variances of all fitted parameters, one 
must account for the contribution from the non-stationary
component of noise as well. This clearly demonstrates the role of the
non-stationary noise component in the fitting procedure. The energy
distribution
amongst the fitted parameters is such that the
non-stationary part of red noise contains as large amount of energy as
the stationary part does (or even more) for the first several parameters. 

An additional point to note is that, if
the number of spin-down parameters is not big enough, the
non-stationary part of the autocovariance function can not be
represented as a
linear combination of products of a smooth function $f(t)$ 
by the fitting function $\psi_a(t)$, and complete filtering out 
non-stationary
component of noise from residuals is impossible. In such a case, if
only the 
stationary
part of the noise is accounted for and the non-stationary part is omitted, 
a negative mean value of residuals can be
obtained that is
physically unadmissible. Taking into account the non-stationary 
noise component brings
the mean value of residuals back to the positive value. By this
consideration,
we would like to underline the role which non-stationary part of noise
plays in
fitting procedure. We consider those models of low frequency noise
which take into account only stationary component to be incomplete and can
lead to
erroneous results. For this reason, we are very careful in dealing
with the
non-stationary noise component in order to properly take into account or
to prove that it is insignificant for calculations. In this context, 
let us note once again that
we have made a shift of the time scale which changes the orbital phase
$u
\mapsto u+\pi {\rm T}$ and introduces additional terms in the
non-stationary
part of the autocovariance function in comparison with expressions for 
$R^{+}(t_{i},t_{j})$ having been given in Kopeikin (1997b).

With these comments we are ready to demonstrate how white and red noise
affect
pulsar timing residuals and covariance matrices of fitting parameters.
   
\subsection{White Phase Noise}

The spectral power of white noise is constant. For this reason, the timing 
residuals are constant as well
\begin{equation}
<r^2>=h_0, \label{wpn1}
\end{equation}
and the covariance matrix of measured parameters $M_{ab}$ coincides 
with the inverse matrix of information (\ref{1.11a}) $L_{ab}^{-1}$. 
The matrix of information $L_{ab}=\frac{m}{2\pi} C_{ab}$, and 
the elements of the inverse matrix $C_{ab}^{-1}$ are given in Tables 
{\ref{tab:2}} - {\ref{tab:5}}.
\begin{table*}
\centering
 \begin{minipage}{120mm}
\centering

  \scriptsize\caption{Elements of matrix of information $C_{ab}$. 
  Quantities ${\rm T}={\pi N}$ and $Q=\cos{\rm T}$.}
  \label{tab:2}
  \begin{tabular}{|c|ccccccc|}
  \hline \\ \\
        &1&3&5&7&9&11&13 \\ 
  \hline \\ \\ 
 1&2{\rm T}&.&.&.&.&.&. \\ \\
 3&$\frac{2{\rm T}^3}{3}$&$\frac{2{\rm T}^5}{5}$&.&.&.&.&. \\ \\
 5&$\frac{2{\rm T}^5}{5}$&$\frac{2{\rm T}^7}{7}$&$\frac{2{\rm T}^9}{9}$&.&.&.&. \\ \\
 7&0&$4{\rm T} Q$&$8{\rm T}^3Q$&${\rm T}$&.&.&. \\ \\
 9&$-2{\rm T} Q$&$-2{\rm T}^3 Q$&$-2{\rm T}^5 Q$&
        $-\frac{{\rm T}}{2}$&$\frac{{\rm T}^3}{2}$&.&. \\ \\
 11&$4{\rm T} Q$&$8{\rm T}^3Q$&$12{\rm T}^5 Q$
     &$\frac{{\rm T}^3}{3}$&$-\frac{{\rm T}^3}{2}$
     &$\frac{{\rm T}^5}{5}$&. \\ \\
 13&$-2{\rm T}^3 Q$&$-2{\rm T}^5 Q$&
     $-2{\rm T}^7 Q$&$-\frac{{\rm T}^3}{2}$&
     $\frac{{\rm T}^5}{5}$&
     $-\frac{{\rm T}^5}{2}$&
     $\frac{{\rm T}^7}{7}$ \\ \\
 \hline 
\end{tabular}
\end{minipage}
\end{table*}\\
\begin{table*}
\centering
\begin{minipage}{120mm}
\centering
\scriptsize\caption{Elements of matrix of information $C_{ab}$. 
Quantities ${\rm T}={\pi N}$ and $Q=\cos{\rm T}$.}
\label{tab:3}
\begin{tabular}{cccccccc}
\hline \\ \\ 
        &2&4&6&8&10&12&14 \\ 
\hline \\ \\ 
 2&$\frac{2{\rm T}^3}{3}$&.&.&.&.&.&. \\ \\
 4&$\frac{2{\rm T}^5}{5}$&$\frac{2{\rm T}^7}{7}$&.&.&.&.&. \\ \\
 6&$\frac{2{\rm T}^7}{7}$&$\frac{2{\rm T}^9}{9}$&$\frac{2{\rm T}^{11}}{11}
 $&.&.&.&. \\ \\
 8&$-2TQ$&$-2{\rm T}^3 Q$&$-2{\rm T}^5Q
 $&$T$&.&.&. \\ \\
 10&$4TQ$&$8{\rm T}^3 Q$&$12{\rm T}^5 Q$&
        $-\frac{T}{2}$&$\frac{{\rm T}^3}{3}$&.&. \\ \\
 12&$-2{\rm T}^3 Q$&$-2{\rm T}^5Q$&$-2{\rm T}^7 
 Q$
     &$\frac{{\rm T}^3}{3}$&$-\frac{{\rm T}^3}{2}$
     &$\frac{{\rm T}^5}{5}$&. \\ \\
 14&$8{\rm T}^3 Q$&$12{\rm T}^5 Q$&
     $16{\rm T}^7 Q$&$-\frac{{\rm T}^3}{2}$&
     $\frac{{\rm T}^5}{5}$&
     $-\frac{{\rm T}^5}{2}$&
     $\frac{{\rm T}^7}{7}$ \\ \\
 \hline 
\end{tabular}
\end{minipage}
\end{table*}
\begin{table*}
\centering
\begin{minipage}{140mm}
\centering
\scriptsize\caption{Elements of the covariance matrix $C_{ab}^{-1}$ of pulsar's 
           parameters for white noise in phase. Quantities ${\rm T}={\pi N}$ and 
           $Q=\cos{\rm T}$.}
  \label{tab:4}         
  \begin{tabular}{cccccccc}
  \hline \\ \\ 
        &1&3&5&7&9&11&13 \\ 
  \hline \\ \\
 1&$\frac{225}{128 {\rm T}}$&.&.&.&.&.&. \\ \\
 3&$-\frac{525}{64 {\rm T}^3}$&$\frac{2205}{32 {\rm T}^5}$&.&.&.&.&. \\ \\
 5&$\frac{945}{128 {\rm T}^5}$&$-\frac{4725}{64 {\rm T}^7}$&$\frac{11025}
     {128 {\rm T}^9}$&.&.&.&. \\ \\
 7&$\frac{315 Q}{8 {\rm T}^3}$&$-\frac{945 Q}{2 
 {\rm T}^5}$&
     $\frac{4725 Q}{8{\rm T}^7}$&$\frac{9}{4 {\rm T}}$&.&.&. \\ \\
 9&$-\frac{225 Q}{16 {\rm T}^3}$&$\frac{1575 Q}{8 
 {\rm T}^5}$&
     $-\frac{4725 Q}{16{\rm T}^7}$&$\frac{45}{8 {\rm T}^3}$&
     $\frac{75}{4 {\rm T}^3}$&.&. \\
     \\
 11&$-\frac{675 Q}{4 {\rm T}^5}$&$\frac{7875 Q}{4 {\rm T}^7}$&
     $-\frac{4725 Q}{2
     {\rm T}^9}$&$-\frac{15}{4 {\rm T}^3}$&$-\frac{225}{8 
     {\rm T}^5}$&$\frac{45}{4 {\rm T}^5}$&. \\ \\
 13&$\frac{525 Q}{16 {\rm T}^5}$&$-\frac{3675 Q}{8 {\rm T}^7}$&
     $\frac{11025 Q}{16
     {\rm T}^9}$&$-\frac{105}{8 {\rm T}^5}$&$-\frac{105}{4 {\rm T}^5}$&
     $\frac{525}
     {8 {\rm T}^7}$&$\frac{175}{4{\rm T}^7} $ \\ \\
 \hline 
\end{tabular}
\end{minipage}
\end{table*}
\begin{table*}
\centering
\begin{minipage}{140mm}
\scriptsize\caption{Elements of the covariance matrix $C_{ab}^{-1}$ of pulsar's 
           parameters for white noise in phase. Quantities ${\rm T}={\pi N}$ and $Q=\cos{\rm T}$.}
\label{tab:5} 
\centering        
  \begin{tabular}{cccccccc}
  \hline\\ \\
    &2&4&6&8&10&12&14 \\ \\ 
  \hline\\ \\ 
 2&$\frac{3675}{128 {\rm T}^3}$&.&.&.&.&.&. \\ \\
 4&$-\frac{6615}{64 {\rm T}^5}$&$\frac{14175}{32 {\rm T}^7}$&.&.&.&.&. \\ \\
 6&$\frac{10395}{128 {\rm T}^7}$&$-\frac{24255}{64 {\rm T}^9}$&$\frac{43659}
     {128 {\rm T}^{11}}$&.&.&.&. \\ \\
 8&$-\frac{315 Q}{16 {\rm T}^3}$&$\frac{945 Q}{8 
 {\rm T}^5}$&
     $-\frac{2079 Q}{16{\rm T}^7}$&$\frac{9}{4 {\rm T}}$&.&.&. \\ \\
 10&$\frac{4725 Q}{2 {\rm T}^5}$&$-\frac{51975 Q}
 {4 {\rm T}^7}$&
     $\frac{51975 Q}{4{\rm T}^9}$&$\frac{45}{8 {\rm T}^3}$&$\frac{75}
     {4 {\rm T}^3}$&.&. \\
     \\
 12&$\frac{1575 Q}{16 {\rm T}^5}$&$-\frac{4725 Q}
   {8 {\rm T}^7}$&
     $\frac{10395 Q}{16{\rm T}^9}$&$-\frac{15}{4 {\rm T}^3}$&
     $-\frac{225}
     {8 {\rm T}^5}$&
     $\frac{45}{4 {\rm T}^5}$&. \\ \\
 14&$-\frac{40425 Q}{8 {\rm T}^7}$&$\frac{55125 Q}
 {2 {\rm T}^9}$&
     $-\frac{218295 Q}{8{\rm T}^{11}}$&$-\frac{105}{8 {\rm T}^5}$
     &$-\frac
     {105}{4 {\rm T}^5}$&
     $\frac{525}{8 {\rm T}^7}$&$\frac{175}{4{\rm T}^7} $ \\ \\
 \hline      
\end{tabular}
\end{minipage}
\end{table*}
On multiplying the main diagonal of $L_{ab}^{-1}$ by the magnitude
of the measurement error $h _0$ yields the variances of the measured
parameters. One can see that precision of the parameter determination depends 
both on
the rate of making observations $m$ and on the number of orbital revolutions
$N$ during the observational session. This is a distinctive feature of
timing measurements in the event when white noise is the only source of errors.
We also note that as the observational span ${\rm T}$ increses the variances 
of all
parameters decrease. Thus, until white noise dominates in observed
timing
residuals, the accuracy of determination of all parameters will be improved.
It is worth emphasizing as well that there is a similarity in the precision of
determination of spin and orbital parameters. Indeed, if one compares variances
for the rotational phase and its time derivatives with corresponding quantities
for orbital motion one finds that dependence on ${\rm T}$ is essentially the
same. For making this statement more obvious we give relationships between
orbital parameters and parameters of the timing model shown in Table 1
\begin{eqnarray}
\label{dop1}
<(\delta\sigma)^2>&=&\frac{1}{x^2}\left(<\beta_7^2>\cos^2\sigma+
<\beta_8^2>\sin^2\sigma\right)\;,\\\nonumber\\
\label{dop2}
\frac{<(\delta n_b)^2>}{n_b^2}&=&\frac{1}{x^2}\left(<\beta_{10}^2>\cos^2\sigma+
<\beta_9^2>\sin^2\sigma\right)\;,\\\nonumber\\ 
\label{dop3}
\frac{<(\delta{\stackrel{.} n_b})^2>}{n_b^2}&=&\frac{4n_b^2}{x^2}
\left(<\beta_{11}^2>\cos^2\sigma+
<\beta_{12}^2>\sin^2\sigma\right)\;,\\\nonumber\\
\label{dop4}
\frac{<(\delta{\stackrel{..} n_b})^2>}{n_b^2}&=&\frac{36n_b^4}{x^2}
\left(<\beta_{14}^2>\cos^2\sigma+
<\beta_{13}^2>\sin^2\sigma\right)\;,\\\nonumber\\
\label{dop5}
<(\delta x)^2>&=&\left(<\beta_8^2>\cos^2\sigma+
<\beta_7^2>\sin^2\sigma\right)\;,\\\nonumber\\
\label{dop6}
<(\delta{\stackrel{.} x})^2>&=&n_b^2\left(<\beta_9^2>\cos^2\sigma+
<\beta_{10}^2>\sin^2\sigma\right)\;,\\\nonumber\\
\label{dop7}
<(\delta{\stackrel{..} x})^2>&=&4n_b^4\left(<\beta_{12}^2>\cos^2\sigma+
<\beta_{11}^2>\sin^2\sigma\right)\;,\\\nonumber\\
\label{dop8}
<(\delta{\stackrel{...} x})^2>&=&36n_b^6\left(<\beta_{13}^2>\cos^2\sigma+
<\beta_{14}^2>\sin^2\sigma\right)\;.
\end{eqnarray}
Calculation of variances of orbital parameters is accomplished using results
shown in Tables \ref{tab:2}-\ref{tab:5}. It is remarkable that there is a
symmetry between variances of odd and even parameters of the timing model, so
that in case of white noise dominance one has, for example,
$<\beta_7^2>=<\beta_8^2>$, and so on. Because of this symmetry there is no
dependence of variances of orbital parameters on the initial orbital phase
$\sigma$. From equations (\ref{dop1})-(\ref{dop8}) we have
\begin{eqnarray}
\label{dop1a}
<(\delta\sigma)^2>&=&\frac{9h_0}{4x^2{\rm T}}\;,\\\nonumber\\
\label{dop2a}
\frac{<(\delta n_b)^2>}{n_b^2}&=&\frac{75h_0}{4x^2{\rm T}^3}\;,\\\nonumber\\ 
\label{dop3a}
\frac{<(\delta{\stackrel{.} n_b})^2>}{n_b^2}&=&\frac{45h_0n_b^2}{x^2{\rm T}^5}
\;,\\\nonumber\\
\label{dop4a}
\frac{<(\delta{\stackrel{..} n_b})^2>}{n_b^2}&=&\frac{1575h_0n_b^4}{x^2{\rm T}^7}
\;,\\\nonumber\\
\label{dop5a}
<(\delta x)^2>&=&\frac{9h_0}{4{\rm T}}\;,\\\nonumber\\
\label{dop6a}
<(\delta{\stackrel{.} x})^2>&=&\frac{75h_0n_b^2}{4{\rm T}^3}\;,\\\nonumber\\
\label{dop7a}
<(\delta{\stackrel{..} x})^2>&=&\frac{45h_0n_b^4}{{\rm T}^5}\;,\\\nonumber\\
\label{dop8a}
<(\delta{\stackrel{...} x})^2>&=&\frac{1575h_0n_b^6}{{\rm T}^7}\;,
\end{eqnarray}
which can be easily compared with variances of corresponding spin-down
parameters in Tables \ref{tab:4}, \ref{tab:5}. In particular, we find that
ratio of root squares of variances of the spin-down parameters to 
corresponding parameters
describing time evolution of orbital phase is approximately equal to $x/P_b$,
that is $<\delta{\cal N}_0/\nu>/<\delta{\cal T}_0>\sim x/P_b$, 
$<\delta\nu/\nu>/<\delta P_b/P_b>\sim x/P_b$, 
$<\delta{\dot\nu}/\nu>/<\delta{\dot P}_b/P_b>\sim x/P_b$, {\it etc.}
As  a concequence of this observation we conclude that in the case of
domination of white noise spin-down parameters of binary pulsars are always
determined better than corresponding parameters of orbital phase.

\subsection{Flicker Noise in Phase ($1/f$ noise)}

Pulsar timing residuals grow with time according to the relationship
\begin{equation}
<r^2>=h _1\left[ \ln(2{\rm T}) -\frac{101}{30}\right] , \label{1.60}
\end{equation}
where $h_1$ is a constant number characterizing intensity of the noise.

The non-stationary part of autocovariance function for PN random walk is
represented as (Kopeikin 1997b):
\begin{equation}
R^{+}(t_i,t_j)=h _1\left[ \ln (u+{\rm T})+\ln (v+{\rm T})\right] ,
 \label{1.61}
\end{equation}
and can be expressed through the basic functions  
\begin{equation}
R^{+}(t_i,t_j)=h _1\left[ \psi _1(v)\ln (u+{\rm T})+\psi
_2(u)\ln (v+{\rm T})\right] . \label{1.62}
\end{equation}
From equation (\ref{1.46}), the contribution from the non-stationary  
part of the flicker noise to the covariance matrix is given by the
integral 
\begin{equation}
M_{ab}^{+}=h_1 \left[\delta_{1a}{\displaystyle \sum_{c=1}^{14}}
L_{bc}^{-1}{\displaystyle \int_{-\rm T}^{\rm T}}\psi_c(u) \ln(u+{\rm
T})du+\delta_{1b}{\displaystyle \sum_{c=1}^{14}}
L_{ac}^{-1}{\displaystyle \int_{-\rm T}^{\rm T}}\psi_c(u) \ln(u+{\rm
T})du \right].
\label{1.63}  
\end{equation}
It is clear from (\ref{1.63}) that the only non-zero elements of the matrix $%
M_{ab}^{+}$ can be $M_{1b}^{+}=M_{b1}^{+}$ 
$(b=1,...,14).$ The most interesting component is the contribution of the 
non-stationary
part to the variance of the first measured parameter
\begin{equation}
M_{11}^{+}=-\frac{361}{240}+2\ln(2{\rm T})
\label{koza}
\end{equation}
\begin{table*}
\begin{minipage}{130mm}
  
\tiny\caption{Elements of the covariance matrix $M_{ab}^{-}$ of pulsar's 
         parameters for flicker noise in phase. Quantities ${\rm T}={\pi N}$, 
         $Q=\cos{\rm T}$, and $\Gamma=\gamma+\ln(2{\rm T})$.
         The
         magnitude of the noise $h_1$ is omitted.}\footnote{Hereafter
         the letter $\gamma=0.577215...$ denotes the Euler's constant.}
  \label{tab:8}
  \tiny        
\begin{tabular}{llllllll}
  \hline\\ \\ 
        &1&3&5&7&9&11&13 \\ 
  \hline\\ \\ 
 1&$\frac{7473}{2560}-\ln(2 {\rm T})$&.&.&.&.&.&. \\ \\
 3&$-\frac{1631}{256 {\rm T}^2}$&$\frac{5145}{128 {\rm T}^4}$&.&.&.&.&. \\ \\
 5&$\frac{2289}{512 {\rm T}^4}$&$-\frac{9555}{256 {\rm T}^6}$&$\frac{19845}
     {512 {\rm T}^8}$&.&.&.&. \\ \\
 7&$\frac{3 Q}{64 {\rm T}^2}[119+60\Gamma]$&
 $-\frac{315 Q}{32 {\rm T}^4}[3+4\Gamma]$&
 $-\frac{315 Q}{64 {\rm T}^4}[1-12\Gamma]$&
 $\frac{9\pi}{4 {\rm T}}$&.&.&. \\ \\
 9&$-\frac{4 Q}{16 {\rm T}^2}$&$\frac{735 Q}{16 
 {\rm T}^4}$&
     $-\frac{945 Q}{16{\rm T}^6}$&$\frac{75\pi}{4 {\rm T}^3}$&
     $\frac{75\pi}{4 {\rm T}^3}$&.&. \\
     \\
 11&$-\frac{ Q}{64 {\rm T}^4}[1273+900\Gamma)]$&
 $-\frac{105 Q}{32 {\rm T}^6}[17+60\Gamma]$&
     $\frac{315 Q}{64 {\rm T}^8}[29-60\Gamma]$&
     $-\frac{15\pi}{4 {\rm T}^3}$&$-\frac{315\pi}{4 {\rm T}^5}$&$\frac{45\pi}
     {4 {\rm T}^5}$&. \\ \\
 13&$\frac{28 Q}{3 {\rm T}^4}$&$-\frac{1715 Q}{16 {\rm
 T}^6}$&
     $\frac{2205 Q}{16
     {\rm T}^8}$&$-\frac{105\pi}{4 {\rm T}^5}$&$-\frac{105\pi}{4 {\rm T}^5}$&
     $\frac{525\pi}
     {4 {\rm T}^7}$&$\frac{175\pi}{4{\rm T}^7} $ \\ \\
 \hline 
\end{tabular}
\end{minipage}
\end{table*}
\begin{table*}
\begin{minipage}{150mm}

\tiny\caption{Elements of the covariance matrix $M_{ab}^{-}$ of pulsar's 
           parameters for flicker noise in phase. Quantities ${\rm T}={\pi N}$, 
         $Q=\cos{\rm T}$, and $\Gamma=\gamma+\ln(2{\rm T})$. The
         magnitude of noise $h_1$ is omitted.}
\label{tab:9}
\tiny
\begin{tabular}{llllllll}
  \hline\\ \\
    &2&4&6&8&10&12&14 \\ \\ 
  \hline \\ \\ 
 2&$\frac{4851}{256{\rm T}^2}$&.&.&.&.&.&. \\ \\
 4&$-\frac{3381}{64 {\rm T}^4}$&$\frac{12495}{64 {\rm T}^6}$&.&.&.&.&. \\ \\
 6&$\frac{46893}{1280 {\rm T}^6}$&$-\frac{4851}{32 {\rm T}^8}$&$\frac{160083}
     {1280 {\rm T}^{10}}$&.&.&.&. \\ \\
 8&$-\frac{693 Q}{160 {\rm T}^2}$&
 $\frac{357 Q}{16 {\rm T}^4}$&
 $-\frac{693 Q}{32 {\rm T}^6}$&
 $\frac{9\pi}{4 {\rm T}}$&.&.&. \\ \\
 10&$\frac{315 Q}{64 {\rm T}^4}[61+20\Gamma]$&
     $-\frac{105 Q}{32 {\rm T}^6}[341+180\Gamma]$&
     $\frac{2079 Q}{64 {\rm T}^8}[21+20\Gamma]$&
     $-\frac{15\pi}{2 {\rm T}^3}$&
     $\frac{75\pi}{4 {\rm T}^3}$&.&. \\
     \\
 12&$\frac{693 Q}{32 {\rm T}^4}$&
 $-\frac{1785 Q}{16 {\rm T}^6}$&
     $\frac{3465 Q}{32 {\rm T}^8}$&
     $-\frac{15\pi}{4 {\rm T}^3}$&$\frac{45\pi}{2 {\rm T}^5}$&$\frac{45\pi}
     {4 {\rm T}^5}$&. \\ \\
 14&$-\frac{147 Q}{64 {\rm T}^6}[261+100\Gamma]$&
     $\frac{735 Q}{32 {\rm T}^8}[91+60\Gamma]$&
     $-\frac{1617 Q}{64 {\rm T}^{10}}[43+60\Gamma]$&
     $-\frac{21 }{2 {\rm T}^6}[289-120\Gamma]$&
     $-\frac{105\pi}{4 {\rm T}^5}$&
     $\frac{315}{ {\rm T}^8}[49-20\Gamma]$&$\frac{175\pi}
     {4{\rm T}^7} $ \\ \\
 \hline
\end{tabular}               
\end{minipage}
\end{table*}
The matrix $M_{ab}^{-}$ is generated by the
stationary part of the autocovariance function and is given in Tables 
{\ref{tab:8}}
and {\ref{tab:9}}. Let us note that, as expected, variances of all measured 
parameters including the first one are positive quantities. 

Comparing results of the variance calculations with those obtained
in previous section for white noise we note that flicker noise in phase worsen
our ability in determining variances of spin-down parameters. At the same time
flicker noise does not disturb variances of orbital parameters and symmetry
between pairs of odd and even parameters. For this reason, using general 
formulae (\ref{dop1})-(\ref{dop8}) one can see that variances of the orbital
parameters do not depend on the initial orbital phase $\sigma$ and have the same
dependence on the total span of observation ${\rm T}$ as it was in case
of white noise. We would like to emphasize the appearance of logarithmic 
terms in timing residuals and the covariance matrices given in Tables 
{\ref{tab:8}} and {\ref{tab:9}}. Logarithmic terms are characteristic for any
flicker noise as it will be shown in subsequent sections. However, practical
observation of the logarithmic terms in temporal behaviour of residuals and
variances of measured parameters is a challenge for observers since it is
difficult to distinguish $\ln{\rm T}$ from constant if observational span is
not long enough. Perhaps, this explains why logarithmic behavior of timing
residuals and variances have not been yet found and flicker noise with spectral
index $s=1$ has not been
ever identified in pulsar timing observations.  
\subsection{Random Walk in Phase ($1/f^2$ noise)}

The timing residuals grow proportionally to the number 
of orbital revolutions $N$%
\begin{equation}
<r^2>=\frac{6}{143}h _2 {\rm T},
\label{1.48c}
\end{equation}
where $h_2$ is a constant number characterizing intensity of the noise.
The non-stationary part of autocovariance function for PN random walk is
represented as (Kopeikin 1997b): 
\begin{equation}
R^{+}(t_i,t_j)  
=h_2{\rm T}\left\{ \psi _1(t_i)\psi _1(t_j)+\frac 1{2{\rm T}}\left[ \psi
_1(t_i)\psi _2(t_j)+\psi _2(t_i)\psi _1(t_j)\right] \right\} . 
\label{1.49}
\end{equation}
Using Eq. (\ref{1.45}) the covariance matrix for non-stationary part of PN
results in
\begin{equation}
\label{1.50}M_{ab}^{+}=h _2{\rm T}\left[ \delta _{a1}\delta _{b1}+
\frac{1}{{2\rm T}}\delta _{a1}\delta _{b2}+
\frac{1}{{2\rm T}}\delta _{b1}\delta _{a2}\right].
\end{equation}
Stationary part of the PN noise gives the covariance matrix displayed in
Tables \ref{tab:10} and \ref{tab:11}.
\begin{table*}
\centering
\begin{minipage}{140mm}
\scriptsize\caption{Elements of the covariance matrix $M_{ab}^{-}$ of pulsar's 
         parameters for random walk in phase. Quantities ${\rm T}={\pi N}$ and $Q=\cos{\rm T}$. The
         magnitude of noise $h_2$ is omitted.}
  \label{tab:10}
  \centering         
  \begin{tabular}{cccccccc}
  \hline\\ \\ 
        &1&3&5&7&9&11&13 \\ 
  \hline\\ \\
 1&$-\frac{45 {\rm T}}{704}$&.&.&.&.&.&. \\ \\
 3&$-\frac{135}{176 {\rm T}}$&$\frac{525}{176 {\rm T}^3}$&.&.&.&.&. \\ \\
 5&$\frac{245}{704 {\rm T}^3}$&$-\frac{105}{44 {\rm T}^5}$&$\frac{1575}
     {704 {\rm T}^7}$&.&.&.&. \\ \\
 7&$\frac{189 Q}{176 {\rm T}}$&
 $-\frac{945 Q}{88 {\rm T}^3}$&
 $-\frac{2205 Q}{176 {\rm T}^5}$&
 $\frac{9}{4 {\rm T}}$&.&.&. \\ \\
 9&$-\frac{45 Q}{176 {\rm T}}$&$\frac{225 Q}{88 {\rm T}^3}$&
     $-\frac{525 Q}{176{\rm T}^5}$&$\frac{15\pi}{88 {\rm T}^3}$&
     $\frac{1275}{44 {\rm T}^3}$&.&. \\
     \\
 11&$-\frac{855 Q}{176 {\rm T}^3}$&
 $\frac{4275 Q}{88 {\rm T}^5}$&
     $-\frac{9975 Q}{176 {\rm T}^7}$&
     $-\frac{15}{4 {\rm T}^3}$&$\frac{765}{88 {\rm T}^5}$&$\frac{45}{4 {\rm T}^5}$&. \\ \\
 13&$\frac{105 Q}{176 {\rm T}^3}$&$-\frac{525 Q}{88 {\rm
 T}^5}$&
     $\frac{1225 Q}{176
     {\rm T}^7}$&$\frac{3045}{88 {\rm T}^5}$&$-\frac{2205}{44 {\rm T}^5}$&
     $-\frac{11025}
     {88 {\rm T}^7}$&$\frac{4375}{44{\rm T}^7} $ \\ \\
 \hline 
\end{tabular}
\end{minipage}
\end{table*}
\begin{table*}
\centering
\begin{minipage}{140mm}
\scriptsize\caption{Elements of the covariance matrix $M_{ab}^{-}$ of pulsar's 
           parameters for random walk in phase. Quantities ${\rm T}={\pi N}$ and $Q=\cos{\rm T}$. The
         magnitude of noise $h_2$ is omitted.}
\label{tab:11}
\centering
\begin{tabular}{cccccccc}
  \hline\\ \\
    &2&4&6&8&10&12&14 \\ \\ 
  \hline \\ \\ 
 2&$\frac{1575}{832{\rm T}}$&.&.&.&.&.&. \\ \\
 4&$-\frac{175}{52 {\rm T}^3}$&$\frac{2205}{208 {\rm T}^5}$&.&.&.&.&. \\ \\
 6&$\frac{1701}{832 {\rm T}^5}$&$-\frac{1575}{208 {\rm T}^7}$&$\frac{4851}
     {832 {\rm T}^9}$&.&.&.&. \\ \\
 8&$-\frac{45 Q}{208 {\rm T}}$&
   $\frac{105 Q}{104 {\rm T}^3}$&
   $-\frac{189 Q}{208 {\rm T}^5}$&
   $\frac{135}{52 {\rm T}}$&.&.&. \\ \\
 10&$\frac{7425 Q}{208 {\rm T}^3}$&
     $-\frac{17325 Q}{104 {\rm T}^5}$&
     $\frac{31185 Q}{208 {\rm T}^7}$&
     $-\frac{6915}{104 {\rm T}^3}$&
     $\frac{75}{4 {\rm T}^3}$&.&. \\
     \\
 12&$\frac{225 Q}{208 {\rm T}^3}$&
    $-\frac{525 Q}{104 {\rm T}^5}$&
     $\frac{945 Q}{208 {\rm T}^7}$&
     $-\frac{285}{52 {\rm T}^3}$&$\frac{31455}{104 {\rm T}^5}$&$\frac{1035}
     {52 {\rm T}^5}$&. \\ \\
 14&$-\frac{16275 Q}{208 {\rm T}^5}$&
     $\frac{37975 Q}{104 {\rm T}^7}$&
     $-\frac{68355 Q}{208 {\rm T}^9}$&
     $\frac{11655 }{104 {\rm T}^5}$&
     $-\frac{105}{4 {\rm T}^5}$&
     $-\frac{58275}{104 {\rm T}^7}$&$\frac{175}{4{\rm T}^7} $ \\ \\
 \hline
\end{tabular}               
\end{minipage}
\end{table*}
One can see that the random walk in phase makes it impossible to track the
rotational phase of the pulsar after amount of orbital revolutions exeeds
$(2816\pi)/(659\nu^2 h_2)\simeq 13.4/(\nu^2h_2)$ since the moment when random walk in 
phase becomes the
dominant source of noise in timing residuals. 
In this case the parameter ${\cal N}_0$ becomes non-informative\footnote{A
parameter is called informative if its mean numerical value obtained in
the fitting procedure is much less than its variance. In the opposite
case, the parameter is called non-informative. We also call the initial
rotational and orbital phases, ${\cal N}_0$ and $\sigma$, informative
parameters if their variances do not exeed $2\pi$.  From a physical point
of view, some of the fitting parameters become non-informative when the
ratio of red timing noise to the deterministic signal (\ref{1.7q}) exeeds a
certain level. The stage at which this happens may be found from
comparision of the mean values of the parameters and their variances.}. 
All other
spin-down parameters can be measured but accuracy of their measurement is lower
comparatively to the cases of white and/or flicker noises in phase. 
On the other hand,
time dependence of variances of orbital parameters is the same as it was in the
case of white noise. Hence, we are still able to determine the initial orbital 
phase $\sigma$, orbital frequency $n_b$, and so on. However, the symmetry between 
pairs of corresponding odd and
even parameters of the timing model under consideration is not preserved. 
For this reason, accuracy in determination of variances of orbital parameters
slightly depends on the initial orbital phase (which remains to be the
informative parameter).       

\subsection{Flicker Noise in Frequency ($1/f^3$ noise)}

The timing residuals are proportional to the square of the
number of orbital revolutions 
\begin{equation}
<r^2>=\frac{1}{105}h _3 {\rm T}^2,
 \label{1.65}
\end{equation}
where $h_3$ is a constant number characterizing intensity of the noise.
The non-stationary part of autocovariance function for flicker noise in
frequency is represented as (Kopeikin 1997b):
\begin{equation}
\label{1.67}
\begin{array}{ll}
R^{+}(t_i,t_j)&  
=\frac{1}{2}h _3{\rm T^2} 
\biggl\{-\psi _1(v)\psi _1(u)-\frac 1{{\rm T}}\psi _1(v)\psi _2(u)+  
-\frac 1{{\rm T}}\psi _2(v)\psi _1(u)-\frac 1{{\rm T^2}}\psi _2(v)\psi
_2(u)+ \\  
\\\mbox{}& 
\left[ \psi _1(v)+\frac 2{{\rm T}}\psi _2(v)+\frac 2{{\rm T}}\psi
_1(v)u-\frac{1}{{\rm T^2}}\psi _1(v)u^2\right] \ln (u+{\rm T})+\\ \\&    
+\left[ \psi _1(u)+\frac 2{{\rm T}}\psi _2(u)+\frac 2{{\rm T}}\psi
_1(u)v-\frac 1{{\rm T^2}}\psi _1(u)v^2\right] \ln (v+{\rm T})\biggr\}. 
\end{array}
\end{equation}
Using Eq. (\ref{1.46}), the covariance matrix for the non-stationary part of the
flicker noise in phase results in 
\begin{equation}
\label{1.68}
\begin{array}{ll}
M_{ab}^{+}&=h _3{\rm T^2} 
\biggl\{-\frac {1}{2}\delta _{a1}\delta _{b1}-
\frac {1}{{2\rm T}}\delta _{a1}\delta_{b2}-
\frac {1}{{2\rm T}}\delta _{b1}\delta_{a2}-
\frac {1}{2{\rm T^2}}\delta _{a2}\delta _{b2}+ \\  \\ 
&\frac {m}{4\pi }
\left[ \delta _{a1}{\displaystyle \sum_{c=1}^{14}}L_{bc}^{-1}
{\displaystyle \int_{-\rm T}^{\rm T}}\psi_c(u)\left( 1-\frac{u^2}{{\rm T^2}}
\right)\ln \left( u+{\rm T}\right) du +
\delta _{b1}{\displaystyle \sum_{c=1}^{14}}L_{ac}^{-1}
{\displaystyle \int_{-\rm T}^{\rm T}}\psi_c(u)\left( 1-\frac{u^2}{{\rm T^2}}
\right)\ln \left( u+{\rm T}\right) du\right]\\  \\ 
&+\frac {m}{2\pi{\rm T}}\left[ \delta _{a2}{\displaystyle \sum_{c=1}^{14}}%
L_{bc}^{-1}{\displaystyle \int_{-\rm T}^{\rm T}} \psi _c(u)\left(
1+\frac {u}{{\rm T}}\right) \ln \left( u+{\rm T}\right) du+
\delta _{b2}{\displaystyle \sum_{c=1}^{14}}%
L_{ac}^{-1}{\displaystyle \int_{-\rm T}^{\rm T}} \psi _c(u)\left(
1+\frac {u}{{\rm T}}\right) \ln \left( u+{\rm T}\right) du
\right]\biggr\} . 
\end{array}
\end{equation}
It is clear from (\ref{1.68}) that the only non-zero elements of the matrix 
$M_{ab}^{+}$ are $M_{1b}^{+}=M_{b1}^{+},$ $%
M_{2b}^{+}=M_{b2}^{+},$ $(b=1,...,14).$ In particular, the elements $M_{11}^{+}$
and $M_{22}^{+}$ which contribute to variances first and second parameters
are given 
\begin{equation}
\label{1.69}
M_{11}^{+}=h _3{\rm T^2}\left[ \ln \left( 2{\rm T}\right) - 
\frac{1991}{1680}\right] , 
\end{equation}
\begin{equation}
\label{1.69a}
M_{22}^{+}=h _3\left[ 2\ln \left( 2{\rm T}\right) +\frac{143 
}{1680}\right] . 
\end{equation}
The stationary part of the covariance matrix $M_{ab}^{-}$ is given in Tables
(\ref{tab:12}) and (\ref{tab:13}). Sum of matrices $M_{ab}^{+}$ and $M_{ab}^{-}$
gives positive numerical values to variances of all measured parameters.
\begin{table*}
\centering
\begin{minipage}{140mm}
\scriptsize\caption{Elements of the covariance matrix $M_{ab}^{-}$ of pulsar's 
         parameters for flicker noise in frequency. Quantities ${\rm T}={\pi N}$ and $Q=\cos{\rm T}$. The
         magnitude of noise $h_3$ is omitted.}
  \label{tab:12}
  \centering         
  \begin{tabular}{lccccccc}
  \hline\\ \\ 
        &1&3&5&7&9&11&13 \\ 
  \hline\\ \\ 
 1&$-\frac{79 \rm T^2}{3584}$&.&.&.&.&.&. \\ \\
 3&$-\frac{5799}{8960}+\frac{1}{2}\ln(2{\rm T})$&$\frac{343}{128 {\rm T}^2}
 $&.&.&.&.&. \\ \\
 5&$\frac{171}{512 {\rm T}^2}$&$-\frac{441}{256 {\rm T}^4}$&$\frac{735}
     {512 {\rm T}^6}$&.&.&.&. \\ \\
 7&$\frac{1553 Q}{2240}$&
 $-\frac{189 Q}{32 {\rm T}^2}$&
 $\frac{399 Q}{64 {\rm T}^4}$&
 $\frac{9\pi}{4 {\rm T}}$&.&.&. \\ \\
 9&$-\frac{79 Q}{448}$&$\frac{49 Q}{32 {\rm T}^2}$&
     $-\frac{105 Q}{64{\rm T}^4}$&$-\frac{12}{\rm T^2}$&
     $\frac{15}{4 {\rm T}^2}$&.&. \\
     \\
 11&$-\frac{1395 Q}{448 {\rm T}^2}$&
 $\frac{847 Q}{32 {\rm T}^4}$&
     $-\frac{1785 Q}{64 {\rm T}^6}$&
     $-\frac{15\pi}{4 {\rm T}^3}$&$\frac{105}{2 {\rm T}^4}$&$\frac{45\pi}{4 {\rm T}^5}$&. \\ \\
 13&$\frac{79 Q}{192 {\rm T}^2}$&$-\frac{343 Q}{96 {\rm T}^4}$&
     $\frac{245 Q}{64
     {\rm T}^6}$&$\frac{28}{{\rm T}^4}$&$-\frac{35}{4 {\rm T}^4}$&
     $-\frac{245}
     {2 {\rm T}^6}$&$\frac{245}{12{\rm T}^6} $ \\ \\
 \hline 
\end{tabular}
\end{minipage}
\end{table*}
\begin{table*}
\centering
\begin{minipage}{140mm}
\scriptsize\caption{Elements of the covariance matrix $M_{ab}^{-}$ of pulsar's 
           parameters for flicker noise in frequency. Quantities ${\rm T}={\pi N}$ and $Q=\cos{\rm T}$. The
         magnitude of noise $h_3$ is omitted.}
\label{tab:13}
\centering
\begin{tabular}{lccccccc}
  \hline\\ \\
    &2&4&6&8&10&12&14 \\ \\ 
  \hline \\ \\ 
 2&$\frac{27241}{17920}-\ln(2{\rm T})$&.&.&.&.&.&. \\ \\
 4&$-\frac{657}{256 {\rm T}^2}$&$\frac{819}{128 {\rm T}^4}$&.&.&.&.&. \\ \\
 6&$\frac{3267}{2560 {\rm T}^4}$&$-\frac{5313}{1280 {\rm T}^6}$&$\frac{7623}
     {2560 {\rm T}^8}$&.&.&.&. \\ \\
 8&$-\frac{4 Q}{35}$&
 $\frac{39 Q}{80 {\rm T}^2}$&
 $-\frac{33 Q}{80 {\rm T}^4}$&
 $\frac{3}{28}$&.&.&. \\ \\
 10&$\frac{3981 Q}{224 {\rm T}^2}$&
     $-\frac{75 Q}{4}$&
     $\frac{2013 Q}{32 {\rm T}^6}$&
     $-\frac{405}{28 {\rm T}^2}$&
     $\frac{75\pi}{4 {\rm T}^3}$&.&. \\
     \\
 12&$\frac{4 Q}{7 {\rm T}^2}$&
 $-\frac{39 Q}{16 {\rm T}^4}$&
     $\frac{33 Q}{16 {\rm T}^6}$&
     $-\frac{15}{28 {\rm T}^2}$&$\frac{2025}{28 {\rm T}^4}$&$\frac{75}
     {28 {\rm T}^4}$&. \\ \\
 14&$-\frac{3725 Q}{96 {\rm T}^4}$&
     $\frac{1309 Q}{8 {\rm T}^6}$&
     $-\frac{4389 Q}{32 {\rm T}^8}$&
     $\frac{125 }{4 {\rm T}^4}$&
     $-\frac{105\pi}{4 {\rm T}^5}$&
     $-\frac{625}{ 4{\rm T}^6}$&$\frac{175\pi}{4{\rm T}^7} $ \\ \\
 \hline
\end{tabular}               
\end{minipage}
\end{table*}

Flicker noise in frequency, if it dominates in timing residuals, will not
improve the accuracy in determination of the initial rotational phase and
frequency. Moreover, variance of the initial rotational phase grows as span of
observations increases. Information about numerical value of the initial 
rotational phase will be completely lost when the root square of the 
variance becomes equal to $2\pi$. Taking into account the non-stationary
component (\ref{1.69}) of the covariance matrix we find that it happens after amount of 
orbital
revolutions $N$ exeeds $\sim 2/(\nu h_3^{1/2})$. For single pulsars, this
corresponds to a time interval of about $\sim 2/(\nu h_3^{1/2})$ years after which the initial rotational
phase becomes the non-informative parameter. We also note that variances of 
the rotational frequency $\nu$ and initial orbital phase $\sigma$ 
can not be determined better than their values obtained before the flicker 
noise in frequency has commenced to corrupt observations. Symmetry in
variances of corresponding pairs of odd and even orbital parameters is completely
destroyed. One interesting consequence of this phenomena is that the
measurement precision of
orbital parameters in the presence of flicker noise  
crucially depends on numerical value of the initial orbital phase until it
remains to be informative parameter. For example, from equations (\ref{dop2}) and    
(\ref{dop2a}) one can see that if we arrange observational data to have
$\sigma\simeq 0$ then parameter ${\dot x}$ is determined better than the
orbital frequency $n_b$. Inversly,
if observational data are prepared to make  $\sigma\simeq \pi/2$ then the
rotational frequency may be measured better than ${\dot x}$. The situation
looks similar to that existing in the quantum mechanical non-demolition 
experiments
(Braginsky \& Khalili 1992) in which the quantum system is prepared in such a 
way
to make the quantum measurement of one of the conjugate variables as good as 
possible at the same time losing information about the other conjugate one.

Certain cases in D'Alessandro {\it et al.} (1997), for which
the derived spectra of the observed timing residuals show spectral
indices $n=3,5$, could well be indicators of the flicker noise
of the respective kind.

\subsection{Random Walk in Frequency ($1/f^4$ noise)}

The timing residuals are proportional to the cube of the
number of orbital revolutions 
\begin{equation}
<r^2>=\frac{2}{6435}h _4 {\rm T}^3, \label{1.52}
\end{equation}
where $h_4$ is the constant quantity characterizing intensity of the noise.
The non-stationary part of the autocovariance function for PN random walk is
represented (Kopeikin 1997b) as  
\begin{equation}
\label{1.53aa}
\begin{array}{ll}
R^{+}(t_i,t_j)& 
=\frac {1}{3}h _4{\rm T^3} 
\biggl\{\psi _1(u)\psi _1(v)+\frac {3}{2{\rm T}}\left[ \psi _1(u)\psi
_2(v)+\psi _2(u)\psi _1(v)\right]  \\  \\ 
&+\frac {3}{{\rm T^2}}\psi _2(u)\psi _2(v)-\frac{1}{4{\rm T^3}}\left[ \psi
_1(u)\psi _4(v)+\psi _4(u)\psi _1(v)\right] +    
+\frac{3}{4{\rm T^3}}\left[ \psi _3(u)\psi _2(v)+\psi _2(u)\psi
_3(v)\right] \biggr\}. 
\end{array}
\end{equation}
Using (\ref{1.45}) the covariance matrix for non-stationary part of PN results
in
\begin{eqnarray}
\label{1.53a}
M_{ab}^{+}&=& 
\frac{1}{3}h _4{\rm T^3}\biggl[\delta _{a1}\delta _{b1}+
\frac {3}{{2\rm T}}\delta _{a1}\delta _{b2}+
\frac {3}{{2\rm T}}\delta _{b1}\delta _{a2}
+\frac{3}{{\rm T^2}}\delta _{a2}\delta _{b2}- 
\frac{1}{4{\rm T^3}}\delta _{a1}\delta _{b4}-
\frac{1}{4{\rm T^3}}\delta _{b1}\delta _{a4}-
\frac 3{4{\rm T^3}}\delta_{a2}\delta _{b3}+
\frac 3{4{\rm T^3}}\delta_{b2}\delta _{a3}\biggr]\;. 
\end{eqnarray}
The stationary part of the noise gives the covariance matrix displayed in Tables
\ref{tab:14} and \ref{tab:15}.
\begin{table*}
\centering
\begin{minipage}{140mm}
\scriptsize
\caption{Elements of the covariance matrix $M_{ab}^{-}$ of pulsar's 
parameters for random walk in frequency. Quantities ${\rm T}={\pi N}$ 
and $Q=\cos{\rm T}$. The
magnitude of noise $h_4$ is omitted.}
  \label{tab:14}
  \centering         
  \begin{tabular}{cccccccc}
  \hline\\ \\ 
        &1&3&5&7&9&11&13 \\ 
  \hline\\ \\ 
 1&$-\frac{95 \rm T^3}{82368}$&.&.&.&.&.&. \\ \\
 3&$\frac{505 {\rm T}}{13728}$&$\frac{805}{2288 {\rm T}}$&.&.&.&.&. \\ \\
 5&$\frac{445}{9152 {\rm T}}$&$-\frac{665}{4576 {\rm T}^3}$&$\frac{945}
     {9152 {\rm T}^5}$&.&.&.&. \\ \\
 7&$\frac{31{\rm T} Q}{572}$&
 $-\frac{219 Q}{572 {\rm T}}$&
 $\frac{105 Q}{286 {\rm T}^3}$&
 $\frac{2493}{572 {\rm T}}$&.&.&. \\ \\
 9&$-\frac{95{\rm T} Q}{6864}$&$\frac{115 Q}{1144{\rm T}}$&
     $-\frac{225 Q}{2288{\rm T}^3}$&$-\frac{15}{26 {\rm T^2}}$&
     $\frac{25}{143 {\rm T}}$&.&. \\
     \\
 11&$-\frac{835 Q}{3432 {\rm T}}$&
 $\frac{245 Q}{143 {\rm T}^3}$&
     $-\frac{1875 Q}{1144 {\rm T}^5}$&
     $-\frac{7515}{572 {\rm T}^3}$&$\frac{725}{286 {\rm T}^3}$&$\frac{30385}{572 {\rm T}^5}$&. \\ \\
 13&$\frac{665 Q}{20592 {\rm T}}$&$-\frac{805 Q}{3432 {\rm
     T}^3}$&
     $\frac{525 Q}{2288
     {\rm T}^5}$&$\frac{35}{26{\rm T}^3}$&$-\frac{175}{429 {\rm T}^3}$&
     $-\frac{5075}
     {858 {\rm T}^5}$&$\frac{1225}{1287{\rm T}^5} $ \\ \\
 \hline 
\end{tabular}
\end{minipage}
\end{table*}\\

\begin{table*}
\centering
\begin{minipage}{140mm}
\scriptsize\caption{Elements of the covariance matrix $M_{ab}^{-}$ of pulsar's 
           parameters for random walk in frequency. Quantities ${\rm T}={\pi N}$ and $Q=\cos{\rm T}$. The
         magnitude of noise $h_4$ is omitted.}
\label{tab:15}
\centering
\begin{tabular}{cccccccc}
  \hline\\ \\
    &2&4&6&8&10&12&14 \\ \\ 
  \hline \\ \\ 
 2&$-\frac{525{\rm T}}{9152}$&.&.&.&.&.&. \\ \\
 4&$-\frac{1295}{4576 {\rm T}}$&$\frac{1015}{2288 {\rm T}^3}$&.&.&.&.&. \\ \\
 6&$\frac{77}{832 {\rm T}^3}$&$-\frac{105}{416 {\rm T}^5}$&$\frac{693}
     {4160 {\rm T}^7}$&.&.&.&. \\ \\
 8&$-\frac{15{\rm T} Q}{2288}$&
 $\frac{29 Q}{1144 {\rm T}}$&
 $-\frac{21 Q}{1040 {\rm T}^3}$&
 $\frac{3{\rm T}}{715}$&.&.&. \\ \\
 10&$\frac{105 Q}{104 {\rm T}}$&
     $-\frac{50 Q}{13{\rm T}^3}$&
     $\frac{315Q}{104{\rm T}^5}$&
     $-\frac{15}{26 {\rm T}}$&
     $\frac{5325}{52 {\rm T}^3}$&.&. \\
     \\
 12&$\frac{75 Q}{2288 {\rm T}}$&
 $-\frac{145 Q}{1144 {\rm T}^3}$&
     $\frac{21 Q}{208 {\rm T}^5}$&
     $-\frac{3}{143 {\rm T}}$&$\frac{75}{26 {\rm T}^3}$&$\frac{15}
     {143 {\rm T}^3}$&. \\ \\
 14&$-\frac{315 Q}{143 {\rm T}^3}$&
     $\frac{4795 Q}{572 {\rm T}^5}$&
     $-\frac{343 Q}{52 {\rm T}^7}$&
     $\frac{357 }{286 {\rm T}^3}$&
     $-\frac{10815}{52 {\rm T}^5}$&
     $-\frac{1785}{ 286{\rm T}^5}$&$\frac{250915}{572{\rm T}^7} $ \\ \\
 \hline
\end{tabular}               
\end{minipage}
\end{table*}
Random walk in frequency significantly contributes to variances of the first three
spin-down parameters so that the initial rotational phase and frequency become
non-informative parameters. Rotational frequency derivative continues to be the
informative parameter but precision of its measurement remains constant and can
not be improved anymore. This has a direct consequence on our ability to use
rotational motion of the pulsar as a time scale (for more detail see Section
8).

Random walk in frequency also converts the initial orbital phase or semi-major
axis of the binary pulsar from the informative to the non-informative 
parameters. Indeed, equations (\ref{dop1}), (\ref{dop5}) and variances of 
parameters
$\beta_7, \beta_8$ taken from Tables \ref{tab:14}, \ref{tab:15} yield 
\begin{eqnarray}
\label{hhh}
<(\delta\sigma)^2>&=&\frac{h_4}{x^2}\left(\frac{2493}{572{\rm T}}\cos^2\sigma+
\frac{3{\rm T}}{715}\sin^2\sigma\right)\;,\\\nonumber\\
\label{ggg}
<(\delta x)^2>&=&h_4\left(\frac{3{\rm T}}{715}\cos^2\sigma+
\frac{2493}{572{\rm T}}\sin^2\sigma\right)\;.
\end{eqnarray}
We can see that if $\sigma$ has been determined prior to the moment
when the random walk in frequency begins to dominate, and its numerical value
was not close to $\pi/2$ or $3\pi/2$, then the information about this value is
lost after $(2860\pi x^2)/(3h_4)\simeq 2995 x^2/h_4$ orbital revolutions. In
the case when $\sigma$ is close to $\pi/2$ or $3\pi/2$ the information will be
lost about parameter $x$. Arranging observations in the interval before random
walk begins a dominant source of the noise we can preserve information either 
about numerical value of $\sigma$ or $x$ parameters. If the initial orbital 
phase becomes the non-informative
parameter one has to average equations (\ref{dop1})-(\ref{dop8}) with respect
to $\sigma$ under assumption that distribution of probability for this
parameter is uniform and concentrated in the interval from $0$ to $2\pi$ (Bard
1974).   

\subsection{Flicker Noise in Frequency Derivative ($1/f^5$ noise)}

The timing residuals are proportional to the forth power of the number of
orbital revolution
\begin{equation}
<r^2>=\frac{1}{8400}h_5 {\rm T}^4. \label{1.71}
\end{equation}
The non-stationary part of the autocovariance function for flicker noise in
frequency derivative 
can be expressed through the basic functions as (Kopeikin 1997b) 
\begin{eqnarray}
\label{1.73}
R^{+}(t_i,t_j)&=&\frac {1}{16}h_5{\rm T^4} 
\biggl\{-5\psi _1(u)\psi _1(v)-\frac{10}{{\rm T}}\left[ \psi _2(u)\psi
_1(v)+\psi _1(u)\psi _2(v)\right] -\\\nonumber\\\mbox{}&& 
\frac{17}{3{\rm T^2}}\left[ \psi _3(u)\psi _1(v)-
\frac{56}{17}\psi
_2(u)\psi _2(v)+\psi _1(u)\psi _3(v)\right]\\\nonumber\\\mbox{}&& 
-\frac{2}{3{\rm T^3}}\left[ \psi _4(u)\psi _1(v)+14\psi _3(u)\psi
_2(v)+14\psi _2(u)\psi _3(v)+\psi _1(u)\psi _4(v)\right]\\\nonumber\\\mbox{}&&   
-\frac{2}{3{\rm T^4}}\left[ \psi _4(u)\psi _2(v)+ 
\frac{11}{2}\psi _3(u)\psi _3(v)+\psi _2(u)\psi _4(v)\right]\biggr\}\\
\nonumber\\\mbox{}&&
+\frac {1}{16}h_5{\rm T^4} 
\biggl\{ 
2\psi _1(v)+\left[\frac{8}{3{\rm T}}u\psi _1(v)+2\psi _2(v)\right]
 +\frac{4}{{\rm T^2}}
\left[2u\psi _2(v)+\psi _3(v)\right] 
+ 
\frac{8}{{\rm T^3}}u\psi _3(v)\\\nonumber\\\mbox{}&&
+\frac{2}{3{\rm T^4}}\left[ u^4\psi
_1(v)-4u^3\psi _2(v)+6u^2\psi _3(v)\right]\biggr\}\ln (u+{\rm T})\\
\nonumber\\\mbox{}&&
+\frac {1}{16}h_5{\rm T^4} 
\biggl\{ 
2\psi _1(u)+\left[\frac{8}{3{\rm T}}v\psi _1(u)+2\psi _2(u)\right]
 +\frac{4}{{\rm T^2}}
\left[2v\psi _2(u)+\psi _3(u)\right] 
+ 
\frac{8}{{\rm T^3}}v\psi _3(u)\\\nonumber\\\mbox{}&&
+\frac{2}{3{\rm T^4}}\left[ v^4\psi
_1(u)-4v^3\psi _2(u)+6v^2\psi _3(u)\right]\biggr\}\ln (v+{\rm T})\;. 
\end{eqnarray}
Using (\ref{1.46}) the covariance matrix for the non-stationary part of the
flicker noise in frequency derivative is reduced to the form 
\begin{eqnarray}
\label{1.74}
M_{ab}^{+}&=&\frac{1}{16}h _5{\rm T^4} 
\biggl[-5\delta_{a1}\delta_{b1}-\frac{10}{{\rm T}}\delta_{a1}\delta_{b2} 
-\frac{10}{{\rm T}}\delta_{b1}\delta_{a2}-
\frac{17}{3{\rm T^2}}\delta _{a1}\delta_{b3}-
\frac{17}{3{\rm T^2}}\delta _{b1}\delta_{a3}+
\frac{56}{3{\rm T^2}}\delta _{a2}\delta _{b2}\\\nonumber\\\mbox{}&&
-\frac{2}{3{\rm T^3}}\delta_{a1}\delta _{b4}
-\frac{2}{3{\rm T^3}}\delta_{b1}\delta _{a4}-
\frac{28}{3{\rm T^3}}\delta _{a2}\delta _{b3}-
\frac{28}{3{\rm T^3}}\delta _{b2}\delta _{a3}
-\frac{2}{3{\rm T^4}}\delta_{a2}\delta _{b4}
-\frac{2}{3{\rm T^4}}\delta_{b2}\delta _{a4}
-\frac{11}{3{\rm T^4}}\delta _{a3}\delta _{b3}\biggr]+ \\
\nonumber\\\mbox{}&&
\frac{m}{2\pi} \biggl\{ \delta_{a1} 
{\displaystyle \sum_{c=1}^{14}}L_{bc}^{-1}
{\displaystyle \int_{-\rm T}^{\rm T}}\psi _c(u)
\left[2{\rm T^4}+\frac{8}{3}{\rm T^3}u+\frac{1}{3}u^4\right]
\ln (u+{\rm T})du+\\\nonumber\\\mbox{}&& 
\delta_{b1} 
{\displaystyle \sum_{c=1}^{14}}L_{ac}^{-1}
{\displaystyle \int_{-\rm T}^{\rm T}}\psi _c(u)
\left[2{\rm T^4}+\frac{8}{3}{\rm T^3}u+\frac{1}{3}u^4\right]
\ln (u+{\rm T})du\biggl\} \\\nonumber\\\mbox{}&& 
+\frac{m}{2\pi} \biggl\{\delta_{a2}{\displaystyle \sum_{c=1}^{14} }L_{bc}^{-1}
{\displaystyle \int_{-\rm T}^{\rm T} }\psi_c(u)\left[
\frac{16}3{\rm T^3}+8{\rm T^2}u
-\frac{8}{3}u^3\right]\ln (u+{\rm T})du+\\ \nonumber\\\mbox{}&&
\delta_{b2}{\displaystyle \sum_{c=1}^{14} }L_{ac}^{-1}
{\displaystyle \int_{-\rm T}^{\rm T} }\psi_c(u)\left[
\frac{16}3{\rm T^3}+8{\rm T^2}u
-\frac{8}{3}u^3\right]\ln (u+{\rm T})du\biggr\}\\\nonumber\\\mbox{}&& 
+\frac{m}{2\pi} \biggl\{\delta_{a3}{\displaystyle \sum_{c=1}^{14} }L_{bc}^{-1}
{\displaystyle \int_{-\rm T}^{\rm T} }\psi_c(u)\left[
4{\rm T^4}+8{\rm T}u+u^2\right]\ln (u+{\rm T})du+\\\nonumber\\\mbox{}&& 
\delta_{b3}{\displaystyle \sum_{c=1}^{14} }L_{ac}^{-1}
{\displaystyle \int_{-\rm T}^{\rm T} }\psi_c(u)\left[
4{\rm T^4}+8{\rm T}u+4u^2\right]\ln (u+{\rm T})du\biggr\}. 
\end{eqnarray}
It is clear from (\ref{1.63}) that the only elements of the matrix $%
M_{ab}^{+}$ being not equal to zero are $M_{1b}^{+}=M_{b1}^{+},$ $%
M_{2b}^{+}=M_{b2}^{+},$ and $M_{3b}^{+}=M_{b3}^{+}$, where the index 
$b=1,...,14.$
In particular,
the elements $M_{11}^{+},M_{22}^{+}$, and $M_{33}^{+}$ are given 
\begin{eqnarray}
\label{1.75}
M_{11}^{+}&=&\frac{1}{4}h_5{\rm T^4}\left[ \ln( 2{\rm T})-\frac{613}{315}
\right]\;,\\
\nonumber\\ 
\label{1.76}
M_{22}^{+}&=&h _5{\rm T^2}\left[ \ln( 2{\rm T}) -\frac{2999}{2520}\right]\;,
\\
\nonumber\\  
\label{1.77}
M_{33}^{+}&=&\frac{1}{2}h_5\left[ \ln( 2{\rm T}) +
\frac{307}{840}\right]\;.
\end{eqnarray}
The matrix $M_{ab}^{-}$ is given in Tables (\ref{tab:16})-(\ref{tab:17}).
\begin{table*}
\centering
\begin{minipage}{140mm}
\scriptsize\caption{Elements of the covariance matrix $M_{ab}^{-}$ of pulsar's 
         parameters for flicker noise in frequency derivative. 
         Quantities ${\rm T}={\pi N}$ and $Q=\cos{\rm T}$. The
         magnitude of noise $h_5$ is omitted.}
  \label{tab:16}
  \centering         
  \begin{tabular}{llcccccc}
  \hline\\ \\ 
        &1&3&5&7&9&11&13 \\ 
  \hline\\ \\ 
 1&$-\frac{13 \rm T^4}{16128}$&.&.&.&.&.&. \\ \\
 3&$\frac{127{\rm T}^2}{8064}$&$\frac{1039}{6720}-\frac{1}{4}\ln(2 {\rm T})$&.&.&.&.&. \\ \\
 5&$\frac{3589}{241920}-\frac{1}{24}\ln(2 {\rm T})$&$-\frac{19}{128{\rm T}^2}$&
     $\frac{21}{256 {\rm T}^4}$&.&.&.&. \\ \\
 7&$\frac{1033{\rm T}^2 Q}{20160}$&
 $-\frac{307 Q}{1120}$&
 $\frac{15 Q}{64 {\rm T}^2}$&
 $\frac{303}{280}$&.&.&. \\ \\
 9&$-\frac{13{\rm T}^2 Q}{1008}$&$\frac{Q}{14}$&
     $-\frac{Q}{16{\rm T}^2}$&$-\frac{17}{56}$&
     $\frac{5}{56}$&.&. \\
     \\
 11&$-\frac{929 Q}{4032}$&
 $\frac{275 Q}{224{\rm T}^2}$&
     $-\frac{67 Q}{64 {\rm T}^4}$&
     $-\frac{269}{56{\rm T}^2}$&$\frac{75}{56 {\rm T}^2}$&$\frac{1195}{56{\rm
     T}^4}$&. \\ \\
 13&$\frac{13 Q}{432}$&$-\frac{Q}{6{\rm
     T}^2}$&
     $\frac{7 Q}{48
     {\rm T}^4}$&$\frac{17}{24{\rm T}^2}$&$-\frac{5}{24{\rm T}^2}$&
     $-\frac{25}
     {8 {\rm T}^4}$&$\frac{35}{72{\rm T}^4} $ \\ \\
 \hline 
\end{tabular}
\end{minipage}
\end{table*}
\begin{table*}
\centering
\begin{minipage}{140mm}
\scriptsize\caption{Elements of the covariance matrix $M_{ab}^{-}$ of pulsar's 
           parameters for flicker noise in frequency derivative. 
           Quantities ${\rm T}={\pi N}$ and $Q=\cos{\rm T}$. The
         magnitude of noise $h_5$ is omitted.}
\label{tab:17}
\centering
\begin{tabular}{lccccccc}
  \hline\\ \\
    &2&4&6&8&10&12&14 \\ \\ 
  \hline \\ \\ 
 2&$-\frac{163{\rm T}^2}{9216}$&.&.&.&.&.&. \\ \\
 4&$-\frac{65657}{483840}+\frac{1}{6}\ln(2{\rm T})$&$\frac{93}{256 {\rm T}^2}
 $&.&.&.&.&. \\ \\
 6&$\frac{1243}{15360 {\rm T}^2}$&$-\frac{429}{2560{\rm T}^4}$&$\frac{2541}
     {25600{\rm T}^6}$&.&.&.&. \\ \\
 8&$-\frac{163{\rm T}^2 Q}{40320}$&
 $\frac{31 Q}{2240}$&
 $-\frac{33Q}{3200{\rm T}^2}$&
 $\frac{{\rm T}^2}{560}$&.&.&. \\ \\
 10&$\frac{2521Q}{4032}$&
     $-\frac{471Q}{224{\rm T}^2}$&
     $\frac{99Q}{64{\rm T}^4}$&
     $-\frac{1}{4}$&
     $\frac{1005}{28{\rm T}^2}$&.&. \\
     \\
 12&$\frac{163Q}{8064}$&
 $-\frac{31Q}{448{\rm T}^2}$&
     $\frac{33Q}{640{\rm T}^4}$&
     $-\frac{1}{112}$&$\frac{5}{4{\rm T}^2}$&$\frac{5}
     {112{\rm T}^2}$&. \\ \\
 14&$-\frac{131Q}{96{\rm T}^2}$&
     $\frac{55Q}{12 {\rm T}^4}$&
     $-\frac{539Q}{160{\rm T}^6}$&
     $\frac{13}{24{\rm T}^2}$&
     $-\frac{935}{12 {\rm T}^4}$&
     $-\frac{65}{ 24{\rm T}^4}$&$\frac{1015}{6{\rm T}^6} $ \\ \\
 \hline
\end{tabular}               
\end{minipage}
\end{table*}
From these tables, equations (\ref{dop1})-(\ref{dop8}), and discussions of
previous sections we conclude that just the flicker noise in frequency
becomes a dominant source of noise in timing residuals - the initial
rotational phase ${\cal N}_0$, rotational frequency $\nu$, the initial orbital
phase $\sigma$, the semi-major axis $x$ - are getting the non-informative 
parameters after
a certain period of time which can be calculated in the same way as it has been
done in previous sections. Determination of numerical values of parameters 
- $\dot{\nu}$, $\dot{x}$, and the orbital frequency $n_b$ can not be improved
comparatively to those values which was obtained at preceding epoch when timing
noise was not so red. This has a consequence for stability of the, so called,
binary pulsar time (BPT) scale (Ilyasov, Kopeikin \& Rodin 1998) which is
discussed in section 8. Flicker noise in frequency may be produced by the
stochastic gravitational wave background. Detailed discussion of the analysis
of timing data from binary pulsars in the presence of this noise 
is given by Kopeikin (1997a). 
\subsection{Random Walk in Frequency Derivative ($1/f^6$ noise)}

The timing residuals are proportional to the fifth power of the number of
orbital revolutions
\begin{equation}
\label{1.55}
<r^2>=\frac{4}{765765}h _6{\rm T^5}, 
\end{equation}
where $h _6$ is a constant quantity characterizing intensity of the noise.
The non-stationary part of the autocovariance function for PN random walk is
represented as (Kopeikin 1997b) 
\begin{eqnarray}
\label{1.57}
R^{+}(t_i,t_j)&=&   
\frac{1}{20}h _6{\rm T^5}\biggl\{\psi _1(u)\psi _1(v)+\frac
{5}{2{\rm T}}\left[ \psi _1(u)\psi _2(v)+\psi _2(u)\psi _1(v)\right] +
\\\nonumber\\\mbox{}&&
+\frac{5}{3{\rm T^2}}\left[ \psi _1(u)\psi _3(v)+4\psi _2(u)\psi
_2(v)+\psi _3(u)\psi _1(v)\right] \\\nonumber\\\mbox{}&&+    
\frac{5}{{\rm T^3}}\left[ \psi _3(u)\psi _2(v)+\psi _2(u)\psi
_3(v)\right] +\frac{5}{{\rm T^4}}\psi _3(u)\psi _3(v) \\\nonumber\\\mbox{}&&   
+\frac{1}{12{\rm T^5}} 
\left[\psi _1(u)\psi _6(v)-5\psi _2(u)\psi _5(v)+10\psi _3(u)\psi
_4(v)\right.\\ \nonumber\\\mbox{}&& 
\left.+10\psi _4(u)\psi _3(v)-5\psi _5(u)\psi _2(v)+\psi _6(u)\psi
_1(v)\right]\biggr\}\;. 
\end{eqnarray}
Using formula (\ref{1.45}) the covariance matrix for non-stationary part of PN 
is reduced to the form 
\begin{eqnarray}
\label{1.58}
M_{ab}^{+}&=& 
\frac{1}{20}h _6{\rm T^5}\biggl[\delta _{a1}\delta _{b1}+
\frac{5}{2{\rm T}}\delta _{a1}\delta _{b2}+
\frac{5}{2{\rm T}}\delta _{b1}\delta _{a2}+
\frac{20}{3{\rm T^2}}\delta _{a2}\delta_{b2}\\\nonumber \\\mbox{}&&+  
\frac{5}{3{\rm T^2}}\delta _{a1}\delta _{b3}+
\frac{5}{3{\rm T^2}}\delta _{b1}\delta _{a3}+
\frac{5}{{\rm T^3}}\delta_{a2}\delta _{b3}+
\frac{5}{{\rm T^3}}\delta_{b2}\delta _{a3}+
\frac{5}{{\rm T^4}}\delta _{a3}\delta _{b3}\\\nonumber \\\mbox{}&&  
+\frac{1}{12{\rm T^5}}\left(\delta_{a1}\delta_{b6}+
\delta_{b1}\delta_{a6}-
5\delta _{a2}\delta_{b5}-5\delta _{b2}\delta_{a5}+
10\delta _{a3}\delta _{b4}+10\delta _{b3}\delta _{a4}\right) \biggr]. 
\end{eqnarray}
The stationary part of the covariance matrix $M_{ab}^{-}$ is given in Tables
(\ref{tab:18})-(\ref{tab:19}).
\begin{table*}
\centering
\begin{minipage}{140mm}
\scriptsize\caption{Elements of the covariance matrix $M_{ab}^{-}$ of pulsar's 
         parameters for random walk in frequency derivative. 
         Quantities ${\rm T}={\pi N}$ and $Q=\cos{\rm T}$. The
         magnitude of noise $h_6$ is omitted.}
  \label{tab:18}
 \centering         
  \begin{tabular}{cccccccc}
  \hline\\ \\ 
        &1&3&5&7&9&11&13 \\ 
  \hline\\ \\ 
 1&$-\frac{43 {\rm T}^5}{480480}$&.&.&.&.&.&. \\ \\
 3&$\frac{53{\rm T}^3}{41184}$&$-\frac{35{\rm T}}{1716}$&.&.&.&.&. \\ \\
 5&$-\frac{3{\rm T}}{572}$&$-\frac{105}{4576{\rm T}}$&
     $\frac{35}{4576{\rm T}^3}$&.&.&.&. \\ \\
 7&$\frac{25{\rm T}^3 Q}{3432}$&
 $-\frac{Q}{44}$&
 $\frac{19Q}{1144 {\rm T}}$&
 $\frac{9{\rm T}}{143}$&.&.&. \\ \\
 9&$-\frac{13{\rm T}^2 Q}{1008}$&$\frac{Q}{14}$&
     $-\frac{Q}{16{\rm T}^2}$&$-\frac{17}{56}$&
     $\frac{5}{56}$&.&. \\
     \\
 11&$-\frac{263{\rm T}Q}{8008}$&
 $\frac{175Q}{1716{\rm T}}$&
     $-\frac{85Q}{1144 {\rm T}^3}$&
     $-\frac{40}{143{\rm T}}$&$\frac{155}{2002{\rm T}}$&$\frac{1245}{1001{\rm
     T}^3}$&. \\ \\
 13&$\frac{43 {\rm T}Q}{10296}$&$-\frac{35Q}{2574{\rm
     T}}$&
     $\frac{35Q}{3432
     {\rm T}^3}$&$\frac{35}{858{\rm T}}$&$-\frac{5}{429{\rm T}}$&
     $-\frac{155}
     {858 {\rm T}^3}$&$\frac{35}{1287{\rm T}^3} $ \\ \\
 \hline 
\end{tabular}
\end{minipage}
\end{table*}
\begin{table*}
\begin{minipage}{140 mm}
\centering
\scriptsize\caption{Elements of the covariance matrix $M_{ab}^{-}$ of pulsar's 
           parameters for random walk in frequency derivative. 
           Quantities ${\rm T}={\pi N}$ and $Q=\cos{\rm T}$. The
         magnitude of noise $h_6$ is omitted.}
\label{tab:19}
\begin{tabular}{cccccccc}
  \hline\\ \\
    &2&4&6&8&10&12&14 \\ \\ 
  \hline \\ \\ 
2&$-\frac{581{\rm T}^3}{700128}$&.&.&.&.&.&. \\ \\
4&$\frac{861{\rm T}}{77792}$&$\frac{105}{2431{\rm T}}$&.&.&.&.&. \\ \\
6&$\frac{189}{17680{\rm T}}$&$-\frac{7}{544{\rm T}^3}$&$\frac{231}
    {35360{\rm T}^5}$&.&.&.&. \\ \\
8&$-\frac{83{\rm T}^3 Q}{291720}$&
 $\frac{2{\rm T}Q}{2431}$&
 $-\frac{Q}{1768{\rm T}}$&
 $\frac{{\rm T}^3}{12155}$&.&.&. \\ \\
 10&$\frac{865{\rm T}Q}{19448}$&
     $-\frac{1225Q}{9724{\rm T}}$&
     $\frac{151Q}{1768{\rm T}^3}$&
     $-\frac{57{\rm T}}{4862}$&
     $\frac{375}{221{\rm T}}$&.&. \\
     \\
12&$\frac{83{\rm T}Q}{58344}$&
 $-\frac{10Q}{2431{\rm T}}$&
     $\frac{5Q}{1768{\rm T}^3}$&
     $-\frac{{\rm T}}{2431}$&$\frac{285}{4862{\rm T}}$&$\frac{5}
    {2431{\rm T}}$&. \\ \\
 14&$-\frac{17003Q}{175032{\rm T}}$&
     $\frac{8015Q}{29172{\rm T}^3}$&
    $-\frac{329Q}{1768{\rm T}^5}$&
     $\frac{371}{14586{\rm T}}$&
    $-\frac{8960}{2431{\rm T}^3}$&
     $-\frac{1855}{14586{\rm T}^3}$&$\frac{1225}{153{\rm T}^5} $ \\ \\
\hline
\end{tabular}               
\end{minipage}
\end{table*}
Random walk in frequency is so strong at low frequencies that time derivatives
of rotational phase also becomes the non-informative parameter after a certain
period of time. Neither the orbital phase, nor the semi-major axis can be
measured precisely. Variances of all parameters which depend on the numerical
value of the initial orbital phase $\sigma$ should be averaged with respect to
$\sigma$ assuming the uniform distribution of probability density function in
the interval from $0$ to $2\pi$. 
\section{Kinematic and Dynamic Pulsar Time Scales}

In this section we discuss stability of time scales based on kinematic rotation
of pulsars (PT scale) and orbital motion of pulsar in a binary system (BPT
scale). This comprehensive analysis can be done due to using results of
calculation of timing residuals and covaraince matrices given in the preceding
sections.
 
The methodology of applying pulsar timing data to fundamental 
metrology and time keeping service was explicitly formulated by Russian 
astronomers in 1979 (Shabanova {\it {\it et al.}} 1979, see also 
Ilyasov {\it {\it et al.}}
1989). Subsequent timing observations definitely 
proved consistency of this approach and made it clear that PT scale 
has a stability comparable with that of
atomic clocks being placed in metrological centres 
(Ilyasov {\it {\it et al.}} 1989, Ginot and
Petit 1991, Kaspi {\it {\it et al.}} 1994). However, it does not mean 
that one should stop looking for other natural periodic physical phenomena 
which might be used for
generating time scales being even more stable than PT. 
A new possibility to
explore the problem is creation and maintenance of modified
ephemeris time (ET) of classical astronomy based not on the orbital motion of
planets in the Solar system (Guinot 1989) but on the orbital motion of pulsar 
in a close 
binary system (Petit and Tavella 1996, Ilyasov {\it et al.} 1998). It has been 
theoretically justified (Damour 1987, Kopeikin 1985,
Sch\"afer 1985) and observationally confirmed with accuracy 0.4\% (Damour and
Taylor 1991, 1992) that orbital motion of pulsar in a binary system is governed by 
laws of General Relativity. For this reason one can predict the orbital motion
of pulsar with extremely high precision on very long time intervals larger than
10 years generating in this way the new time scale BPT. However, 
the principal question arises about the limit on the stability of such
time scale in the presence of red stochastic noise process in pulsar timing
residuals. We answer this question in the present section.
\subsection{Binary pulsars as time keeping standards}
Standards of time and frequency have three important characteristics:
(1) frequency stability, (2) reproduction of the time scale unit, and 
(3) the span of life time. Regarding these characteristics we note that the 
life time of binary pulsars reaches $10^6$ years and more which makes them very
long-lived time standards. Orbital motion of pulsar in a close binary system is
practically not subjected to external gravitational perturbations. 
For this reason the
orbital motion has a very high intrinsic stability allowing reproduction of
duration of one orbital revolution to very high precision. 
Binary pulsar time scale - BPT (Petit \& Tavella 1996, Ilyasov {\it et al.} 
1998) can be constructed according to the formula          
\begin{eqnarray}
\label{11}
{\cal T}&=&{\cal T}_0+P_b N+\frac {1}{2}{\dot P}_b N^2,
\end{eqnarray}
where ${\cal T}_0$ is an initial epoch for orbital phase, $N$ is the 
number of orbital revolutions, $P_b$ and ${\dot P}_b$ are the orbital period 
and its time derivative reffered
to the initial epoch ${\cal T}_0$. This equation does not 
contain terms
depending on high-order time derivatives of the orbital period as they  
equal identically to
zero in the barycentric reference frame of the binary system
(Kopeikin 1985, Kopeikin and Potapov 1994). However, radial acceleration of the
binary system with respect to observer and its proper motion
in the sky bring about appearance of the second and next high-order time 
derivatives both of $P_b$ (Damour \& Taylor 1992, Bell \& Bailes 1996) and 
of the projected semi-major axis of the pulsar's orbit 
(Kopeikin 1996, 1997a)(see equation (\ref{1.7q}) for more detail). 
These effects change the intrinsic values of $P_b$ and 
${\dot P}_b$ and make more difficult their determination from observations.

An additional serious problem arises with the possible presence in TOA 
the low-frequency noise of astrophysical origin. Indeed, intrinsic rotational 
motion of pulsar may be perturbed by the random walk in phase and/or its time 
derivatives (Cordes \& Greenstein 1981), interstellar medium and relic 
stochastic gravitational 
waves change randomly propagation of electromagnetic pulses from the pulsar 
to observer being the reason for contamination of timing residuals by flicker 
noise (Mashhoon \& Grishchuk 1980, Bertotti {\it et al.} 1983,
Blandford {\it et al.} 1984, Backer \& Foster 1990, Kopeikin 1997a). 
As a result, timing residuals
represent a pure deterministic process upon which the stohastic additive 
noise is imposed. This noise consists of mixture of white noise of
errors of observations and the low-frequency red noise of astrophysical origin.
The
spectrum $S(f)$ of the red noise is described in Table \ref{tab:1}. 
It contains
both stationary and non-stationary components (Kopeikin 1997b). The presence of
the red noise makes specific restrictions on the stability of both PT and BPT
time scales which we briefly discuss in what follows. More detail on the 
subject may be found in the paper by Ilyasov, Kopeikin \& Rodin (1998).

\subsection{BPT scale and its stability}

A complete treatment of the problem under consideration requires rather
combersome calculations. In order to avoid mathematical difficulties we
restrict ourselves to a simplified model of a binary system with the pulsar on
the circular orbit so that all results of previous sections can be applied
without any restrictions. Our goal is to get the optimal estimate of measured
stability of BPT on the background of additive, low-frequency
red noise in TOA. This problem had been partially considered (see, for
instance, Van Trees 1968,
Tikhonov 1983). Unfortunately, the methods developed in most previous 
publications can not be
applied directly in our study so that we need to rely upon our own approach. 
In the problem under consideration the signal
represents a linear combination of rotational phase of the pulsar, which is
given by a
polynomial of time, and quasi-periodic sinusoidal function with the argument
being the orbital phase which is also a polynomial of time. As we have assumed
the noise is additive to the signal, we can use the functional expression
for the analyzed signal in the form:  
\begin{eqnarray} \label{eq1}
\xi (t)&=&{\cal N}(t,\beta_a )+\nu\;\epsilon(t),\quad 0\le t\le \tau\;,
\end{eqnarray}
where ${\cal N}(t,\beta_a )$ is function of time describing deterministic 
part
of the signal, depending on measured parameters $\beta_a $, and given by
equation (\ref{1.7q}); $\nu$ is the pulsar rotational frequency; 
$\epsilon(t)$ is a random process defined by equation (\ref{1.7aa}) which
differentiate between
noises which are true instabilities of the pulsar clock, and apparent
instabilities that arise from random galactic accelerations, gravitational
waves, etc. The autocorrelation function of the noise process $\epsilon(t)$ 
is given in
Table \ref{tab:1}, $\tau$ is the total span of observational time.  
We also assume that observations are evenly spaced along the orbit with
frequency $m$. It has allowed us to replace in all sums by integrals
and to treat the stochastic process $\epsilon(t)$ as continuous.
\begin{figure}\centering
\unitlength=0.7pt
\begin{picture}(493.00,661.00)(100.00,160.00)
\thicklines
\put(104.00,480.00){\line(2,-5){48.00}}
\put(152.00,360.00){\line(6,-5){55.00}}
\put(207.00,314.00){\line(4,-1){72.00}}
\put(279.00,296.00){\line(6,1){81.00}}
\put(360.00,310.00){\line(5,2){61.00}}
\put(421.00,335.00){\line(6,5){53.00}}
\put(474.00,380.00){\line(1,2){36.00}}
\put(104.00,773.00){\line(2,-5){177.00}}
\put(282.00,329.00){\line(6,-5){74.00}}
\put(356.00,268.00){\line(4,-1){64.00}}
\put(420.00,252.00){\line(1,0){52.00}}
\put(473.00,252.00){\line(6,5){52.00}}
\thinlines
\put(152.00,359.00){\line(0,-1){199.00}}
\put(206.00,315.00){\line(0,-1){155.00}}
\put(281.00,329.00){\line(0,-1){169.00}}
\put(359.00,309.00){\line(0,-1){149.00}}
\put(420.00,333.00){\line(0,-1){173.00}}
\put(473.00,379.00){\line(0,-1){219.00}}
\thicklines
\put(100.00,160.00){\vector(1,0){434.00}}
\put(100.00,160.00){\vector(0,1){640.00}}
\put(100.00,145.00){\makebox(0,0)[cc]{0}}
\put(152.00,145.00){\makebox(0,0)[cc]{$\tau_1$}}
\put(206.00,145.00){\makebox(0,0)[cc]{$\tau_2$}}
\put(280.00,145.00){\makebox(0,0)[cc]{$\tau_3$}}
\put(359.00,145.00){\makebox(0,0)[cc]{$\tau_4$}}
\put(420.00,145.00){\makebox(0,0)[cc]{$\tau_5$}}
\put(473.00,145.00){\makebox(0,0)[cc]{$\tau_6$}}
\put(528.00,145.00){\makebox(0,0)[cc]{$
\tau$}}
\put(71.00,794.00){\makebox(0,0)[cc]{$
\sigma(\tau)$}}
\put(129.00,773.00){\makebox(0,0)[cc]{$\sigma_v(\tau)$}}
\put(129.00,480.00){\makebox(0,0)[cc]{$\sigma_y(\tau)$}}
\put(120.00,390.00){\makebox(0,0)[cc]{$\tau^{-3/2}$}}
\put(171.00,325.00){\makebox(0,0)[cc]{$\tau^{-1}$}}
\put(237.00,295.00){\makebox(0,0)[cc]{$\tau^{-1/2}$}}
\put(330.00,325.00){\makebox(0,0)[cc]{$(\ln\tau)^{1/2}$}}
\put(391.00,340.00){\makebox(0,0)[cc]{$\tau^{1/2}$}}
\put(464.00,345.00){\makebox(0,0)[cc]{$\tau(\ln\tau)^{1/2}$}}
\put(510.00,411.00){\makebox(0,0)[cc]{$\tau^{3/2}$}}
\put(510.00,265.00){\makebox(0,0)[cc]{$\tau^{1/2}$}}
\put(380.00,245.00){\makebox(0,0)[cc]{$\tau^{-1/2}$}}
\put(199.00,474.00){\makebox(0,0)[cc]{$\tau^{-3/2}$}}
\put(320.00,280.00){\makebox(0,0)[cc]{$\tau^{-1}$}}
\put(445.00,240.00){\makebox(0,0)[cc]{{\rm const}}}
\end{picture}
\vspace{0.5cm}
\caption{Characteristic behavior of variances characterizing stability of 
intrinsic rotational frequency of pulsar $\sigma_y$ and that of the orbital 
frequency of the binary system $\sigma_v$. In the case of evenly spaced
observations theoretically predicted ratio 
$\sigma_y/\sigma_v$ is equal to $1.75\pi\; x/P_b\simeq 10^{-3}\div 10^{-4}$ 
until white noise dominates in timing residuals.   
Displacement of two curves in subsequent moments of time
depends on the relative magnitude of white and red noise components.}
\label{fig1}
\end{figure}
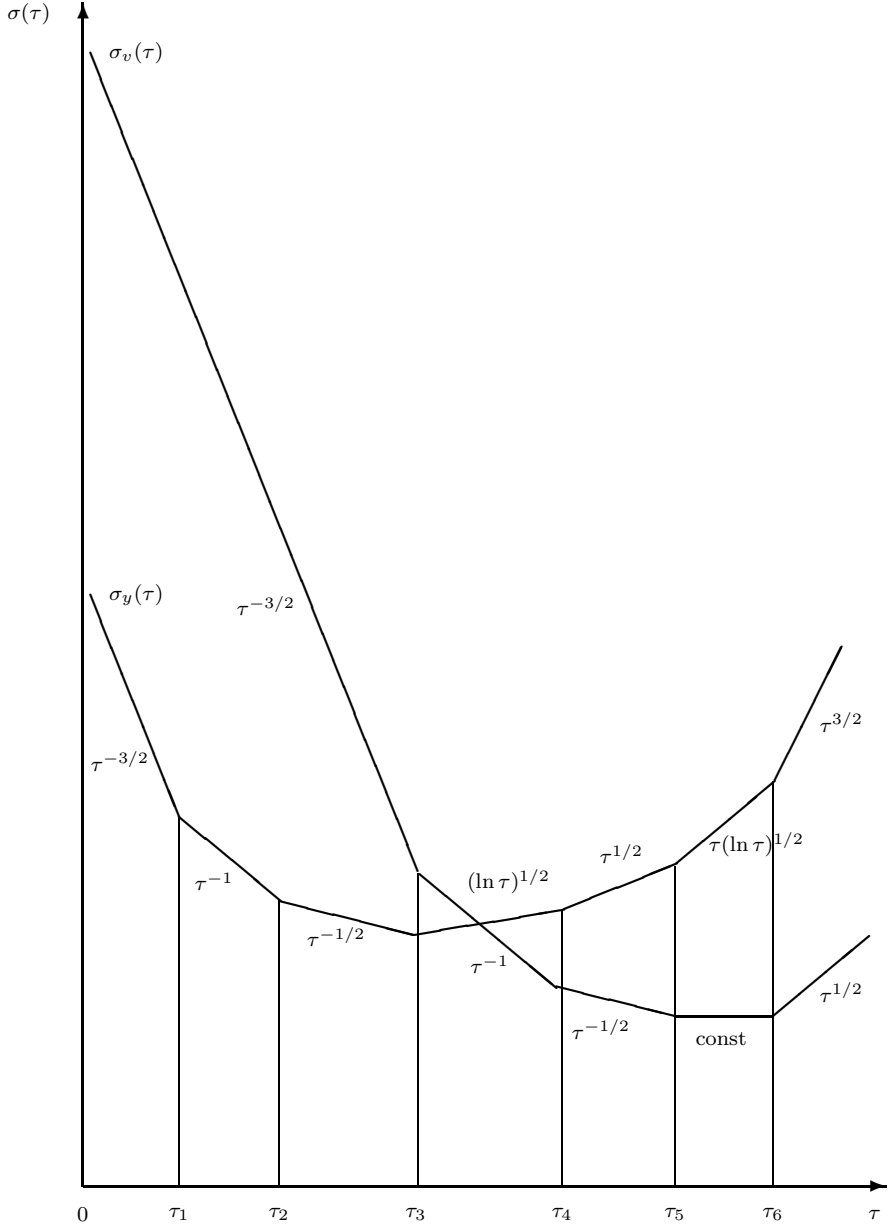

For an adequate treatment of stability of time scales PT and BPT it is
convenient to introduce two new parameters $y=\delta\nu/\nu$ and 
$v=\delta n_b/n_b$, where $\delta\nu$ and $\delta n_b$
represent differences between true physical value of the parameter in question
and its estimate obtained by the least squares method when 
fitting TOA of the pulsar. Variances of estimates of rotational and orbital
frequencies ($\sigma_y^2$, and, $\sigma_v^2$, respectively)
are convenient for comparing  stability of two time scales - dynamic
BPT and kinematic PT. Mathematical expressions for these variances are 
called
true variances (Rutman 1978) and represent the close analogue of the Allan
variance being used in fundamental metrology for analysing stability of 
time-frequency standards. Expressions for $\sigma_v$ and $\sigma_y$ crucially 
depend on the type of noise present and can be evaluated from 
the variance of $\beta_2$ and from 
equation (\ref{dop2}) and variances $\beta_7$, $\beta_8$. Their explicit
dependence on the amplitude of noise $h_s$,
$s=0,\;1,\;2,\ldots,\;6$ and duration of the total span of observations $\tau$ 
are
given in Table \ref{tab:sup1} and are ploted schematically in Fig. \ref{fig1} 
where the horizontal
axis shows the total span of observations $\tau$ and vertical axis
numerates values of the parameters $\sigma_y$ and $\sigma_v$. It is worth 
emphasizing that $\sigma_y$ can not depend on orbital parameters of the pulsar 
because it characterizes instability of the
intrinsic rotational frequency of the pulsar and, for this reason, 
can not be related with its orbital motion. On the other hand,
the quantity $\sigma_v$ naturally depends on the orbital period of the binary
$P_b=2\pi/n_b$ and its projected semi-major axis $x$ since
these parameters characterize frequency and amplitude of the orbital 
sine wave in timing residuals. Additional functional dependence of $\sigma_v$ 
on the initial orbital phase $\sigma$ in the case of random walk in phase 
is explained by the fact that the orbital
frequency of the binary is one of the quadrature component of the sine function
(the second quadrature component is the first time derivative of the projected
semi-major axis $\dot{x}$ as follows from the definition of parameters
$\beta_9$ and $\beta_{10}$ given in Table 1).
Low-frequency red noise with spectral index $s=2$ distroys an equilibrium in
the accuracy of simultaneous determination of the quadrature components 
and leads to the dependence of $\sigma_v$ on the parameter $\sigma$. In the
case of red noise with index $s\ge 3$ the initial orbital phase is the
non-informative parameter. Thus, we have averaged the variances with
respect to $\sigma$ which, for this reason, does not appear in the
corresponding expressions for $\sigma_v$.  

In constructing Fig. \ref{fig1} we have assumed that during the time interval 
$[\tau_0,\tau_1]$ white phase noise dominates. This noise is caused by the 
presence of 
errors of measurements in TOA and, possibly, by the uncorrelated noise in the 
intrinsic rotation of the pulsar. Various components of the 
low-frequency red noise commence
to appear on time intervals $\tau > \tau_1$. Their amplitudes $h_s$, as the
experience of working with atomic clocks shows, gradually decrease as the 
spectral index $s$ of the noise increases. For instance, in the model 
being considered in Fig. \ref{fig1} the red flicker noise with spectral density 
$S(f)\simeq h_1/f$ 
dominates in timing residuals from the moment $\tau_1$ till $\tau_2$. Then, 
starting from the moment $\tau_2$ 
up to $\tau_3$ the random walk in phase with the spectral density 
$S(f)\simeq h_2/f^2$
dominates, and so on.  In general, we assume that the longer time interval of
observations the more significant the contribution of the red noise with 
higher
numerical value of the spectral index $s$. Such behavior relates to the fact 
that usually 
the red noise with higher spectral index has lower intensity
(that is, smaller amplitude $h_s$) and can give rise to a noticable 
contribution only on longer time interval $\tau$ from the beginning of
observation.

It is interesting to note that behavior of $\sigma_y (\tau)$ is reminiscent 
of the, so-called, narrow-band frequency dispersion
used in the fundamental metrology for characterizing stability of time-frequency 
standards when low-frequency noise is dominating (Rutman, 1978). However, in
the timing model under consideration, $\sigma_y(\tau)$, turns out to be
significantly dependent on the non-stationary part of red noise with the
spectral indices $s=3$ or high (see Table \ref{tab:sup1}). As there is no
satisfactory theoretical model of the 
non-stationary component such dependence of $\sigma_y(\tau)$ is an indication
of necessity of working out the other, more practical approach for
estimation of the variance characterizing stability of intrinsic rotational
motion of pulsar and PT scale. It seems that the modified measure of the 
variance, which is called $\sigma_z$ instead of $\sigma_y$, introduced by 
Taylor (1992) and based on
the using of mathematical technique of transfer filter functions in spectral
frequency domain is rather constructive and fruitful step ahead toward this
direction. In particular, Matsakis {\it et al.} (1997) have recently derived 
the
explicit expression for $\sigma_z(\tau)$ which is proportional to the second
time derivative of pulsar's rotational frequency $\dot{\nu}$. We have used 
our analytic estimation of pulsar parameters in timing model (\ref{1.7q})
in order to calculate $\sigma_z$. Results of the calculations is shown in 
Table \ref{tab:sup1} along with $\sigma_y$ and $\sigma_v$. We confirm the
statement of Matsakis {\it et al.} (1997) that 
function $\sigma_z$ is independent from the
non-stationary component of red noise. Moreover, it does not contain
slow logarithmic trends explicitly appearing in $\sigma_y$ for flicker noises
with indices $s=3$ and $s=5$. 

Comparing the $\sigma_z$ statistic with 
$\sigma_v$ we find that $\sigma_v$ is not sensitive to
the non-stationary component of red noise as well as $\sigma_z$, and does not 
contain the
logarifmic trends. However, $\sigma_v$ may depend on the initial orbital phase
for red noises with indices $s \ge 2$ if span of observational time is not long
enough to make that parameter to be non-informative. Our calculations also 
clearly show that measured value of the variance $\sigma_v(\tau)$
is not sensitive to red noise with index $s\le 2$ and, for this reason, does
not allow one to distingiush such red noise from the white one. Nevertheless,
this
specific  behavior of $\sigma_v(\tau)$ opens a possibility of using
orbital motion of the pulsar as a new time reference being more stable than PT
scale on longer intervals.
\begin{table*}
\begin{minipage}{140mm}
\centering
\caption{Dependence of variances $\sigma_y(\tau)$, $\sigma_z(\tau)$, and 
$\sigma_v(\tau)$ on the total span of observations $\tau$ for red noise with
different spectra. Herein,
$h_s,\;(s=1,2\ldots,6)$ is the amplitude of the corresponding noise component; 
$P_b$ is the orbital period and $x$ is the projected semi-major axis of the
binary orbit, $\sigma$ is the initial orbital phase,
$\Delta t$ is the interval of time between two successive observations,
$C_3,...,C_6$ are positive constants of order 0.05 - 0.5 reflecting dependence
of the variances on the non-stationary component of the noise model.}
\label{tab:sup1}
\begin{tabular}{cccc}
\hline
&&&\\
$S(f)$ & $\qquad\sigma_y^2(\tau)$ & $\qquad\sigma_z^2(\tau)$
& $\qquad\sigma_v^2(\tau)$\\&&&\\
\hline
&&&\\
$h_0$ &$\frac{3675}{16}\Delta t\,h_0\,\tau^{-3}$&
$\frac{2835}{16}\Delta t\,{h_0}\,\tau^{-3}$&
$\frac{75}{2\pi^2}\frac{P_b^2}{x^2}\Delta t\,h_0\,\tau^{-3}$ \\&&&\\
$h_1/f$ &$\frac{4851}{64}{h_1}{\tau^{-2}}$&
$\frac{2499}{64}{h_1}{\tau^{-2}}$&
$\frac{75}{4\pi^2}\frac{P_b^3}{x^2}h_1\tau^{-3}$\\&&&\\
$h_2/f^2$ &$\frac{1575}{416}{h_2}{\tau^{-1}}$ &
$\frac{441}{416}{h_2}{\tau^{-1}}$ &
$\frac{1275}{88\pi^4}\frac{P_b^4}{x^2}(\sin^2\sigma+\frac{11}{17}\cos^2
\sigma)h_2\tau^{-3}$\\&&&\\
$h_3/f^3$ &$(C_3+\ln\tau)h_3$&
$\frac{819}{2560}{h_3}$&
$\frac{15}{32\pi^4}\frac{P_b^4}{x^2}\,h_3\tau^{-2}$\\ &&&\\
$h_4/f^4$ &$(C_4-\frac{525}{18304})h_4\tau$&
$\frac{203}{18304}{h_4}\tau$ &
$\frac{25}{2288\pi^4}\frac{P_b^4}{x^2}\,h_4\tau^{-1}$\\ &&&\\
$h_5/f^5$ &$\frac14(C_5+\ln\tau)h_5\tau^2$&
$\frac{93}{20480}{h_5}\tau^2$ &
$\frac{5}{1792\pi^4}\frac{P_b^4}{x^2}\,h_5$\\ &&&\\
$h_6/f^6$ &$(C_6-\frac{581}{5601024}){h_6}\tau^3$&
$\frac{21}{77792}{h_6}\tau^3$&
$\frac{5}{64064\pi^4}\frac{P_b^4}{x^2}\,h_6\tau$\\ &&&\\
\hline
\end{tabular}
\end{minipage}
\end{table*}
Indeed, as one can see in Fig. \ref{fig1}, function $\sigma_y(\tau)$\footnote{
Behavior of $\sigma_z$ is more representative for characterizing stability of
PT scale. However, graphic behavior of $\sigma_y$ looks approximately the 
same as that of $\sigma_z$.} 
describing stability of PT scale begins to grow from the moment $\tau_3$ but
$\sigma_v(\tau)$ characterizing stability of BPT scale
continues to decrease until the red noise with spectral index $s \ge 5$ begins 
to
dominate. This relative behavior of two variances is quite general and 
does not depend on the specific
numerical value of intensities of red noise. Theoretical analysis reveals, 
that the minimum of $\sigma_v(\tau)$  
is reached significantly later than that of $\sigma_y(\tau)$ 
(and $\sigma_z(\tau)$ respectively). Depth of the
minimum for $\sigma_y(\tau)$ (and $\sigma_z(\tau)$ depends on the noise with 
spectral index $s=3$ 
which is produced by the large-scale inhomogeneities of the interstellar medium 
(Blandford {\it et al.} 1984). Depth of the minimum for $\sigma_v(\tau)$ 
depends on
the amplitude of noises with spectral indecies $s= 5$ which exist due to the
stohastic background of relic gravitational radiation produced by
physical processes in early universe (Mashhoon \& Grishchuk 1980,
Bertotti {\it et al.} 1983, Kopeikin 1997a). In principle, noise with spectral index $s=5$ may 
arise
also as a result of random fluctuations of the first time derivative of
intrinsic rotational frequency of the pulsar (Cordes \& Greenstein 1981).
The existence of such fluctuations is extremely unlikely and, for this reason,
they are
not discussed in the following discussion. As an example we specify the
numerical value of the minimum of $\sigma_z(\tau)$ for a single
millisecond pulsar PSR B1937+21 which reach $10^{-14}$ on time interval 2-3
years. Existence of the minimum with subsequent turning-up of the curve 
$\sigma_z(\tau)$ for this pulsar is explained by the dominance of
random instabilities in the rotational phase of the pulsar (Kaspi {\it et al.} 
1994).

It is remarkable from principal and practical points of view that depth of the
absolute minimum for the function, $\sigma_v(\tau)$, can be predicted with a 
high degree of
certainty. It is defined by the amplitude of the relic stochastic
gravitational wave background radiation with spectral index $s=5$. More
specifically, for binary pulsars with circular orbits the depth of minimum of 
$\sigma_v(\tau)$ is defined by the expression:
\begin{equation}\label{fon}
\sigma_v\simeq 1.2\cdot 10^{-20}\sqrt{\Omega_g}P_b^2x^{-1}{\rm h},
\end{equation}
where $\Omega_g$ is energy density of stochastic gravitational wave background
radiation per logarithmic unit interval of frequency, $P_b$ is the orbital
period of the binary measured in seconds, $x$ is the projected semi-major axis 
of the pulsar's orbit measured in seconds, ${\rm h}$ is the Hubble 
constant in units 100 km c$^{-1}$Mpc$^{-1}$. It is worthwhile to point out 
(as the formula (\ref{fon}) shows) 
that numerical value of the minimum for $\sigma_v(\tau)$ depends not only on 
the
magnitude of fundamental cosmological parameters $\Omega_g$ and ${\rm h}$ but 
on the
numerical values of orbital parameters of the binary system as well. Of course,
for binary pulsars with elliptical orbits formula (5) will also include
dependence on the eccentricity of the orbit. We shall consider this more
complicated case elsewhere. 

Formula (\ref{fon}) yields the absolute (fundamental) value of the minimum for 
$\sigma_v(\tau)$.
However, the real minimum of $\sigma_v(\tau)$ depends significantly on the
amplitude of noise with spectral index $s=6$. If the amplitude of this noise 
is less than that of the stochastic noise of gravitational wave backround 
radiation, then the real minimum of $\sigma_v(\tau)$ coincides with absolute
minimum given by formula (\ref{fon}). In the opposite case the minimum of
$\sigma_v(\tau)$ will be located above the absolute minimum (\ref{fon}). In any
case, the observed numerical value of minimum for $\sigma_v(\tau)$ may, in
principle, reach the level $10^{-15}\div 10^{-16}$ under assumption that 
$\Omega_g h^2\le 10^{-8}$ (Starobinsky 1979, Vilenkin 1981, 
Rubakov {\it et al.} 1982).
The result obtained can be used in two ways. First, existence of fundamental
minimum for $\sigma_v(\tau)$ allows to search for (or to establish the upper
limit on) the stochastic gravitational wave radiation by means of
using binary pulsars with long orbital periods and large ratio of $P_b/x$ which
is necessary for reducing the absolute value of minimum of $\sigma_v(\tau)$ as 
much as possible. That will reduce the interval of time $\tau$ 
being required in order to reach the minimum. Second,
using binary pulsars with short orbital periods and small ratio of 
$P_b/x$ one can make the minimum of $\sigma_v(\tau)$ as deep as possible in
order to make the interval of stability of BPT scale as long as possible.
We argue that determination of numerical value of minimum of 
$\sigma_v(\tau)$ from observations may give very likely more precise indicator 
of
the upper limit on the energy density of stochastic gravitational radiation 
than
measuring the slope of function $\sigma_z(\tau)$. The matter is that
determination of minimum of $\sigma_v(\tau)$ is more statistically reliable
than one of the slope of $\sigma_z(\tau)$ which depends to a large extent on
uncertainty in calculation of errors of measurement of the curve on
long time intervals (Kaspi {\it et al.} 1994).

\section{Conclusions}

We have done analytic calculations of time dependence of residuals and 
covariance matrix of fitting
parameters of a binary pulsar under assumption that observations are evenly
spaced. The results have been used to estimate the observed stability of 
intrinsic 
rotational and orbital frequencies of the pulsar. It was shown
that the dynamic time scale BPT may be more stable under certain circumstances
than the kinematic PT scale. It was clarified that stability of BPT is 
restricted (like that of PT) by the presence of red noise even when the orbital 
motion of the pulsar can be considered as pure
deterministic. Nevertheless, this restriction on stability of BPT is important 
only if observations are run on time intervals being significantly longer than 
characteristic time of 
instability of PT. This remarkable property of BPT leads us to hope that its
usage may help in future to undertake deeper analysis of laws of
gravitational physics. There is no doubt that long-term stability of BPT time 
scale makes it as a useful practical tool for fundamental metrology of time.    

Investigation of problem of construction and maintanace of BPT scale for
studing various problems of modern astronomy relates tightly to the problem of
identification of nature and spectral characteristic of timing noise presenting
in TOA. There is a hope that publication of papers of 
Deshpande {\it et al.} (1996) and D'Alessandro {\it et al.} (1997) 
devoted to detailed discussion of timing noise in the case of a sample of 
18 single pulsars will attract a new interest to this problem. In particular, a
reader of these papers will find a handful cases where the slopes are possibly 
more consistent with the spectral indices $n=3, 5$ etc. than say $n=4, 6$.
Of course, the uncertainty in some of these cases is large but still there is 
an indicative support for the flicker noise cases. 

Theoretical development of ideas relating to the study of stability of BPT time
scale can be continued in several directions. Probably, one should
consider binary pulsars with elliptical orbits having maximal ratio $P_b/x$
which can improve quality of the time scale and, consequently, testing General
Relativity in the strong field regime. In this problem it will be necessary to 
take into account
possible variations of orbital elements of the binary caused by different, not
well predicted factors, like flybys of stars or gravitational waves with
frequncies being close to the orbital one (Chicone {\it et al.} 1998). 
It is not so difficult to show that for
the BPT it is preferable to take binary pulsars with small
enough ratio $m_p/m_c$ ($m_p$ and $m_c$ are masses of pulsar and its companion
respectively), long semi-major axis of the pulsar's orbit $a_p$, short orbital
period $P_b$, and $\sin i\simeq 1$, that is large $x=a_p\sin i/c$. 
It would be desirable and
worthwhile to create a network of reference binary pulsars and start their
continuous timing observations in leading radio astronomical observatories 
of the world in order to make the concept of BPT time scale practically 
available as soon as possible.\\ \\
{\bf ACKNOWLEDGEMENTS}
 
I am grateful to N. Wex, A.V. Gusev, and Yu.P. Ilyasov
for fruitful discussions and 
R. Rieth for assistance in preparation of the manuscript. I thank D. Lorimer
for careful reading the manuscript and numerous valuable comments and 
A. A. Deshpande for useful remarks. It is a
pleasure to acknowledge the hospitality of G. Neugebauer and G. Sch\"afer and
other members of the Institute for Theoretical Physics of the Friedrich
Schiller University of Jena. 
This work was
supported by the Th\"uringer Ministerium f\"ur Wissenschaft, Forschung und 
Kultur grant No B501-96060.

\label{lastpage}

\begin{thebibliography}{99}
\bibitem{} Backer D. C., Foster R. S., 1990, ApJ, {\bf 361}, 300
\bibitem{b44} Bard Y., 1974, Nonlinear Parameter Estimation, Academic Press:
    New York
\bibitem{b42} Bell J.F., Bailes M., 1996, {\bf 456}, L33
\bibitem{b22} Bertotti B., Carr B.J., Rees M.J., 1983, MNRAS, {\bf 203}, 945
\bibitem{b23} Blandford R., Narayan R., Romani R., 1984, J. Astrophys. Astron.,
    {\bf 5}, 369
\bibitem{}    Braginsky V. B., Khalili F. Ya., 1992, Quantum Measurement,
    Cambridge University Press, Cambridge
\bibitem{b20} Brumberg V.A., Kopeikin S.M., 1989, Nuov. Cim., {\bf B103}, 63
\bibitem{b14} Brumberg V.A., Kopeikin S.M., 1990, Cel. Mech. Dyn. Astron., {\bf 
48}, 23
\bibitem{} Chikone C., Mashhoon B., Retzloff D.G., 1998, Chaos in the Kepler
System, preprint gr-qc/9806107, 26 June 1998
\bibitem{b26} Cordes J.M., 1978, ApJ, {\bf 222}, 1006
\bibitem{b26'} Cordes J.M., 1980, ApJ, {\bf 237}, 216
\bibitem{b27} Cordes J.M., Helfand D.J., 1980, ApJ, {\bf 239}, 640 
\bibitem{b28} Cordes J.M., Greenstein, G., 1981, ApJ, {\bf 245}, 1060 
\bibitem{b29} Cordes J.M., Downs G.S.,  1985, ApJ Suppl., {\bf 59}, 343 
\bibitem{} D'Alessandro F., Deshpande A.A., McCulloch P.M., 1997, 
J.Astrophys. \& Astr., {\bf 18}, 5
\bibitem{b38} Damour T., 1975, Nuovo Cim., {\bf B26}, 157
\bibitem{b7} Damour T., 1983a, Phys. Rev. Lett., {\bf 51}, 1019
\bibitem{b39} Damour T., 1983b, In: Gravitational Radiation, eds. 
     N. Deruelle and T. Piran, North-Holland: Amsterdam, p. 59
\bibitem{} Damour T., 1987, In: 300 Years of Gravitation, Eds. S.W. Hawking and
W. Israel, Cambridge: Cambridge University Press, p. 128
\bibitem{b2} Damour T., Sch\"afer G., 1988, Nuov. Cim., {\bf B101}, 127     
\bibitem{b1} Damour T, Grishchuk L.P., Kopeikin S.M., Sch\"afer G., 1989, 
  Higher-order
  Relativistic Dynamics of Binary Systems, In: Proc. 5th MG Meeting on General
  Relativity, eds. D.G. Blair \& M.J. Buckingham, World Scientific: 
  Singapore, p.451  
\bibitem{b41} Damour T., Taylor J.H., 1991, ApJ, {\bf 366}, 501
\bibitem{} Damour T., Taylor J. H., 1992, Phys. Rev. D, {\bf 45}, 1840.
\bibitem{b45} Davis P.J., Rabinowitz P., 1984, Methods of Numerical Integration
    Academic Press: New York
\bibitem{b30} Deeter J.E., Boynton P.E.,  1982, ApJ, {\bf 261}, 337
\bibitem{b31} Deeter J.E., 1984, ApJ, {\bf 281}, 482
\bibitem{} Deeter, J. E., Boynton P.E., Lamb, F. K., Zylstra, G., 1989, ApJ,
{\bf 336}, 376
\bibitem{} Deshpande, A. A., D'Alessandro, F., McCulloch, P. M.,
        1996,  J. Astrophys. \& Astr., {\bf 17}, 7
\bibitem{b18a} Doroshenko O.V., Kopeikin S.M., 1990, Sov. Astron., {\bf 34}, 496 
\bibitem{b18} Doroshenko O.V., Kopeikin S. M., 1995, MNRAS, {\bf 274}, 1029     
\bibitem{b15} Fukushima T., 1995, A\&A, {\bf 294}, 895
\bibitem{b37} Gel'fand I.M., Shilov G.E., 1964, Generilized Functions:
    Properties and Operations, Academic Press, New York  
\bibitem{b8} Grishchuk L.P., Kopeikin S.M., 1983, Sov. Astron. Lett., {\bf 9}, 
230
\bibitem{b25} Groth E.J., 1975, ApJ Suppl., {\bf 29}, 453
\bibitem{} Guinot B., 1989, In: Reference Frames, eds. J.Kovalevsky {\it et al},
Kluwer: Dordrecht, p. 351
\bibitem{} Guinot B., Petit G., 1991, Astron. \& Astrophys., {\bf 248}, 292
\bibitem{b24} Hogan C.J., Rees M.J., 1984, Nature (London), {\bf 311}, 109
\bibitem{Hooge} Hooge F.N., 1976, Physica, {\bf 83B}, 14
\bibitem{ht} Hulse R.A., Taylor J.H., 1975, ApJ, {\bf 195}, L51
\bibitem{b11} Ilyasov Yu.P., Kuzmin A.D., Shabanova T.V., Shitov Yu.P., 1989,
   Proc. FIAN, {\bf 199}, 149
\bibitem{}    Ilyasov Yu.P., Kopeikin S. M., Rodin A. E., 1998, Astronomy
Letters, {\bf 24}, 228
\bibitem{b47a} Iyer B. R., Will C. M., 1995, Phys. Rev. D, {\bf 52}, 6882 
\bibitem{jar} Jaranowski P., 1997, in: Mathematics of Gravitation. Part II.
Gravitational Wave Detection, ed. A. Kr\'olak, Banach Center Publications of
Inst. Math. Polish Acad. Sci., {\bf 41}, 55
\bibitem{b47b} Jaranowski P., Sch\"afer G., 1997, Phys. Rev. D, {\bf 55}, 4712
\bibitem{b12} Kaspi V.M., Taylor J.H., Ryba M.F., 1994, ApJ, {\bf 428}, 713 
\bibitem{} Kopeikin S. M., 1985, Sov. Astron., {\bf 62}, 889 
\bibitem{b19} Kopeikin S.M., 1988, Cel. Mech., {\bf 44}, 87
\bibitem{} Kopeikin S.M., 1994, ApJ, {\bf 434}, L67
\bibitem{b43'} Kopeikin S.M., 1996, ApJ, {\bf 467}, L93
\bibitem{b36} Kopeikin S.M., 1997a, Phys. Rev. D, {\bf 56}, 4455
\bibitem{b35} Kopeikin S.M., 1997b, MNRAS, {\bf 288}, 129
\bibitem{b5} Kopeikin S.M., Potapov, V.A., 1994, Astron. Reports, {\bf 38}, 104
\bibitem{lor} Lorimer D., 1996, Physics World, {\bf 9}, No 2, 25
\bibitem{} Mashhoon, B., Grishchuk, L.P., 1980, ApJ, {\bf 236}, 990
\bibitem{} Matsakis D. N., Taylor J. H., Eubanks T. M., 1997, Astron.
\& Astrophys., {\bf 326}, 924
\bibitem{b34} McHugh M.P., Zalamansky G., Vernotte F., Lantz E., 1996, Phys.
    Rev. D, {\bf 54}, 5993
\bibitem{b3} Ohta T., Kimura T., 1988, Phys. Lett. A, {\bf 129}, 436
\bibitem{pm} Peters P. C., Mathews J., 1963, Phys. Rev., {\bf 131}, 435
\bibitem{p} Peters P. C., 1964, Phys. Rev., {\bf 136}, 1224
\bibitem{} Petit G., Tavella P., 1996, Astron. \& Astrophys., {\bf 308}, 290
\bibitem{rbj} Rickett B. J., 1990, Ann. Rev. Astron. Astrophys., {\bf 28}, 
561     
\bibitem{b17} Rickett B. J., 1996, Interstellar Scaterring: Observations and
  Interpretations, In: Pulsars, Problems and Progress, eds. S. Johnston, M.A.
  Walker, M. Bailes, ASP Conf. Ser. {\bf 105}, 439
\bibitem{Planat} Planat M., Goirdano V., Marianneau G., Vernotte F., Mourey M.,
Eckert C., Mieh\'e J. A., 1996, IEEE Trans. on Ultrasonics and Freq. Control,
{\bf 43}, 326   
\bibitem{b47} Rieth R., 1997 (work in progress) 
\bibitem{} Rubakov V., Sazhin M. V., Veryaskin A., 1982, Phys. Lett., {\bf
115B}, 189 
\bibitem{} Rutman J., 1978, Proc. IEEE, {\bf 66}, 1048 
\bibitem{b40} Sch\"afer G., 1985, Ann. Phys. N.Y., {\bf 161}, 81  
\bibitem{b4} Sch\"afer G., Wex N., 1993, Phys. Lett. A, {\bf 174}, 196 ; 
Erratum: Phys. Lett. A, {\bf 177}, 461, (1993)
\bibitem{b6} Sch\"afer G., 1995, The general relativistic two-body problem, In:
   Symposia Gaussiana, eds. Behara/Fritsch/Lintz, Walter de Gruyter \& Co: 
   Berlin, p. 667
\bibitem{} Shabanova T.V., Il'in V.G., Ilyasov Yu.P., Ivanova Yu.D., 
Kuz'min A.D., Palii G.N., Shitov Yu.P., 1979, Measuring Technique,
No 10, p. 73 (in Russian)
\bibitem{b16} Shapiro I.I., 1964, Phys. Rev. Lett., {\bf 13}, 789   
\bibitem{b13} Standish E.M., 1990, A\&A, {\bf 233}, 252
\bibitem{} Starobinsky A., 1979, Sov. Phys. - JETP Lett., {\bf 30}, 682  
\bibitem{b32} Stinebring D.R., Ryba M.F., Taylor J.H., Romani R.W., 1989, Phys.
    Rev. Lett., {\bf 65}, 285 
\bibitem{} Stratonovich R. L., Sosulin, Yu. G., 1966, Radiotekhnika i
Elektronika, {\bf 11}, 497 (in Russian)  
\bibitem{b9} Taylor J.H., Weisberg J.M., 1982, ApJ, {\bf 253}, 908 
\bibitem{b10} Taylor J.H., Weisberg J.M., 1989, ApJ, {\bf 345}, 434 
\bibitem{b33} Thorsett S.E., Dewey R.J., Phys. Rev. D, {\bf 53}, 3468
\bibitem{} Tikhonov V.I., 1983, Optimal Detection of Signals, Radio and 
Communication: Moscow (in Russian)
\bibitem{Van Trees} Van Trees H. L., 1968, Detection, Estimation, and Modulation
Theory, John Willey \& Sons: New York
\bibitem{} Vilenkin A., 1981, Phys. Rev. D, {\bf 24}, 2082 
\bibitem{ww} Wex N., 1995, Class. Quantum Grav., {\bf 12}, 983
\bibitem{b46} Wex N., 1997, privite communication
\end{thebibliography}
\end{document}